# Carbon Nanotubes in Biology and Medicine: *in vitro* and *in vivo* Detection, Imaging and Drug Delivery

Zhuang Liu, Scott Tabakman, Kevin Welsher, Hongjie Dai*

Department of Chemistry, Stanford University, CA 94305

hdai@stanford.edu

**ABSTRACT**

Carbon nanotubes exhibit many unique intrinsic physical and chemical properties and have been intensively explored for biological and biomedical applications in the past few years. In this comprehensive review, we summarize the main results of our and other groups in this field and clarify that surface functionalization is critical to the behaviors of carbon nanotubes in biological systems. Ultra-sensitive detection of biological species with carbon nanotubes can be realized after surface passivation to inhibit the non-specific binding of bio-molecules on the hydrophobic nanotube surface. Electrical nanosensors based on nanotubes provide a label-free approach to biological detections. Surface enhanced Raman spectroscopy of carbon nanotubes opens up a method of protein microarray with down to 1 fM detection sensitivity. *In vitro* and *in vivo* toxicity studies reveal that well water soluble and serum stable nanotubes are biocompatible, non-toxic and potentially useful for biomedical applications. *In vivo* biodistributions vary with the functionalization and possibly also sizes of nanotubes, with a tendency of accumulation in the reticuloendothelial systems (RES), including the liver and spleen, after intravenous administration. If well functionalized, nanotubes may be excreted mainly through the biliary pathway in feces. Carbon nanotube-based drug delivery has shown promises in various *in vitro* and *in vivo* experiments including delivery pf small interfering RNA (siRNA), paclitaxel and doxorubicin. Moreover, single-walled carbon nanotubes with various interesting intrinsic optical properties have been used as novel photoluminance, Raman and photoacoustic contrast agents for imaging of cells and animals. Further multidisciplinary explorations in this field are promising and may bring new opportunities to the realm of biomedicine.

# 1. Introduction

Nanomaterials have sizes ranging from about one nanometer up to several hundred nanometers, comparable to many biological macromolecules such as enzymes, antibodies, DNA plasmids, etc. Materials in this size range exhibit interesting physical properties, distinct from both the molecular and bulk scales, presenting new opportunities for biomedical research and applications in various areas including biology and medicine. As an emerging field, nanobiotechnology bridges the physical sciences with biological sciences via chemical methods in developing novel tools and platforms for understanding biological systems and disease diagnosis and treatment [1-3].

Carbon nanotubes (CNTs) are rolled up seamless cylinders of graphene sheets, exhibiting many unique physical, mechanical and chemical properties which have gleaned tremendous interest in the past decade [4-8]. Depending on the number of graphene layers from which a single nanotube is composed, CNTs are classified as single-walled carbon nanotubes (SWNTs) and multi-walled carbon nanotubes (MWNTs). Applications of CNTs span many fields and applications, including composite materials [9], nano-electronics [10, 11], field-effect emitters [12], and hydrogen storage [13]. In recent years, efforts have also been devoted to exploring the potential biological applications of CNTs as motivated by their interesting size, shape, and structure, as well as attractive, unique physical properties [14-17].

With diameters of 1~2 nm, and length ranging from as short as 50 nm up to 1 cm, SWNTs are one dimensional (1-D) nanomaterials which may behave distinctly from spherical nanoparticles in biological environments, offering new opportunities in biomedical research. The flexible 1-D nanotube may bend to facilitate multiple binding sites of a functionalized nanotube to one cell, leading to a multi-valence effect, and improved affinity of nanotubes conjugated with targeting ligands. With all atoms exposed on the surface, SWNTs have ultra-high surface area (theoretically 1300$m^2$/g) that permits efficient loading of multiple molecules along the length of the nanotube sidewall. Moreover, supramolecular binding of aromatic molecules can be easily achieved by π-π stacking of those molecules onto the poly-aromatic surface of nanotubes [18].

SWNTs are quasi 1-D quantum wires with sharp densities of electronic states (electronic DOS) at the van Hove singularities (Fig. 1a), which impart unique optical properties to SWNTs [19]. SWNTs are highly absorbing materials with strong optical absorption in the NIR range due to E11 optical transitions (Fig. 1a&b), and thus can be utilized for photothermal therapy [20, 21] and photoacoustic imaging [22].

Semiconducting SWNTs with small band gaps on the order of 1 eV exhibit photoluminescence in the near infrared (NIR) range. The emission range of SWNTs is 800 nm – 2000 nm [17, 23, 24], which covers the biological tissue transparency window, and is therefore suitable for biological imaging. SWNTs also have unique resonance-enhanced Raman signatures for Raman detection / imaging, with large scattering cross-sections for single tubes [25, 26]. The intrinsic physical properties of SWNTs can be utilized for multimodality imaging and therapy.

Different from SWNTs, MWNTs are formed by multiple layers of graphene and have much larger diameters (10-100 nm). Although MWNTs exhibit less rich and attractive optical properties than SWNTs, their use in biological systems could be different from that of SWNTs due to their larger sizes, which could offer different platforms for different purposes, such as delivery of large biomolecules including DNA plasmids into cells [27-30].

Motivated by various properties of CNTs, research towards applying carbon nanotubes for biomedical applications has been progressing rapidly. CNT-based sensors have been developed to detect biological species including proteins and DNA [14, 31, 32]. Relying on their optical properties, SWNTs can be utilized as optical tags or contrast agents for various biological imaging techniques [17, 22, 24, 26]. Others, in addition to our group, have uncovered that properly functionalized CNTs are able to enter cells without toxicity, shuttling various biological molecular cargoes into cells [15, 16, 29, 33-36]. Our latest study has shown promise of using CNTs for *in vivo* cancer treatment in a mouse model [37]. However, despite those exciting findings, researchers have reported the negative sides of CNTs, showing that non-functionalized nanotubes are toxic to cells and animals [38-43]. The biodistribution and long-term fate of CNTs have been explored by us and several different groups, obtaining different results from different methods and materials [44-50]. These controversial findings require clarification to avoid confusion to the public.

In this review, we first review various routes used to functionalize carbon nanotubes including covalent and non-covalent methods. Carbon nanotube-based electronic and optical biosensors are then discussed. Surveying our and others' results, we next summarize that while non-functionalized, hydrophobic CNTs have shown toxicity [38-43], those with carefully designed biocompatible coatings are harmless to cells *in vitro* [17, 18, 20, 51-57] and *in vivo* at least to mice within tested dose ranges [45, 58]. In terms of biodistribution, although direct comparison between different studies may not be fair because of the different CNT materials used, tracking SWNTs themselves by their intrinsic physical properties (Raman scattering, photoluminescence, $^{13}$C isotope mass spectrum) shows that, similar to other nanoparticles *in vivo*, SWNTs after systemic administration are dominantly localized in reticuloendothelial systems (RES) including the liver and spleen [45, 48, 50]. Moreover, we review the current progress of using carbon nanotubes for *in vitro* drug delivery studies as well as pioneering efforts towards *in vivo* cancer treatment. Lastly, the SWNT based biomedical imaging *in vitro* and *in vivo* are discussed.

## 2. Functionalization of carbon nanotubes for biological applications

As grown, raw carbon nanotubes have highly hydrophobic surfaces, and are not soluble in aqueous solutions. Surface chemistry or functionalization is required to solubilize CNTs, and to render biocompatibility and low toxicity for biomedical applications. Surface functionalization of carbon nanotubes may be covalent or non-covalent. Chemical reactions forming bonds with nanotube sidewalls are carried out in the covalent functionalization case, while non-covalent functionalization exploits favorable interactions between the hydrophobic domain of an amphiphilic molecule and the CNT surface, affording aqueous nanotubes wrapped by surfactant.

**2.1 Covalent functionalization of carbon nanotubes**

Various covalent reactions have been developed to functionalize carbon nanotubes. Among them oxidization is one of the most common. CNT oxidization is carried out with oxidizing agents such as nitric acid [59, 60]. During the process, carboxyl groups are formed at the ends of tubes as well as at the defects on the sidewalls. Zeng et al. observed $sp^3$ carbons on SWNTs after oxidization and further covalent conjugation with amino acid [61]. However, although oxidized CNTs are soluble in water, they will aggregate in the presence of salt due to charge screening effects, and thus cannot be directly used for biological applications due to the high salt content of most biological solutions. Further modification can be achieved by attaching oxidized CNTs with hydrophilic polymers such as poly(ethylene glycol) (PEG), yielding CNT-polymer conjugates stable in biological environments (Fig. 2a) [18, 58, 62]. We have used covalently PEGylated SWNTs synthesized by this strategy for in vitro and in vivo applications [18, 58].

Another type of widely used type of covalent reaction to functionalize CNTs is the cylcoaddition reaction, which occurs on the aromatic sidewall, instead of nanotube ends and defects as in the oxidization case. [2+1] cycloadditions can be conducted by reacting CNTs with azides through photochemistry (Fig. 2b) [63, 64] or carbene generating compounds via the Bingel reaction (Fig. 2c) [65, 66]. A 1,3-dipolar cycloaddition reaction on CNTs developed by Prato et al. is now a commonly used reaction (Fig. 2d) [67, 68]. Azomethine-ylide generated by condensation of an α-amino acid and an aldehyde is added to the graphitic surface, forming a pyrrolidine ring coupled to the CNT sidewall. Functional groups (e.g. amino-terminated PEG) introduced from the modified α-amino acid can be used for further conjugation of biological molecules such as peptides or drugs [36, 69].

Despite the robustness of the covalent functionalization method, the intrinsic physical properties of CNTs such as photoluminescence and Raman scattering are often destroyed after chemical reactions due to the disrupted nanotube structure. The intensities of Raman scattering and photoluminescence of SWNTs are drastically decreased after covalent modification, reducing the potential of optical applications of this material.

**2.2 Non-covalent functionalization of carbon nanotubes**

In contrast to covalent functionalization, non-covalent functionalization of CNTs can be carried out by coating CNTs with amphiphilic surfactant molecules or polymers. Since the chemical structure of the π-network of carbon nanotubes is not disrupted, except for shortening of length due to sonication employed in the functionalization process, the physical properties of CNTs are essentially preserved by non-covalent approach. Consequently, aqueous solutions of CNTs, especially SWNTs, engineered by non-covalent functionalization are promising for multiple biomedical applications including imaging.

The poly-aromatic graphitic surface of a carbon nanotube is accessible to the binding of aromatic molecules via π-π stacking [70, 71]. Taking advantage of the π-π interaction between pyrene and the nanotube surface, we and others have used pyrene derivatives to non-covalently functionalize carbon nanotubes (Fig. 3a) [70, 72]. Chen et al. showed that proteins can be immobilized on SWNTs functionalized by an amine reactive pyrene derivative [70]. A recent study conducted by Wu et al. also used pyrene conjugated glycodendrimers to solublize carbon nanotubes [72]. Beside pyrene derivatives, single stranded DNA molecules have been widely used to solubilize SWNTs by the π-π stacking between aromatic DNA base units and the nanotube surface (Fig. 3b) [20, 73, 74]. However, a recent report by Moon et al. showed that DNA molecules coated on SWNTs could be cleaved by nucleases in the serum, suggesting that DNA functionalization of SWNTs might not be stable in biological environments with nucleases [75]. We also have shown that fluorescein (FITC) terminated PEG chains are able to solubilize SWNTs with the aromatic FITC domain π-π stacked on the nanotube surface, yielding visibly fluorescent SWNTs useful for biological detection and imaging [76]. Furthermore, other aromatic molecules such as porphyrin derivatives have also been used for non-covalent functionalization of CNTs [77].

Various amphiphiles have been used to suspend carbon nanotubes in aqueous solutions, with hydrophobic domains attached to the nanotube surface via van der Waals forces and hydrophobic effects, and polar heads for water solubility [78]. We used tween-20 and pluronic triblock copolymer to non-covalently functionalize nanotube surfaces to reduce the non-specific binding of proteins in the case of SWNT-based biosensors [14]. Pluronic tri-block polymer was used by Cherukuri et al. to solubilize SWNTs for in vivo experiments [50]. However, the pluronic coating is not sufficiently stable and is quickly replaced by serum proteins once SWNTs are intravenously injected. Other traditional surfactants including sodium dodecyl sulfate (SDS) and Triton X-100 have also been used to suspend CNTs in water [79]. Carbon nanotubes solubilized by those amphiphiles with relatively high critical micelle concentrations (CMC) are typically not stable without an excess of surfactant molecules in the solution. Large amounts of surfactants may lyse cell membranes and denature proteins, and are therefore not useful in biological environments.

An ideal non-covalent functionlization coating on CNTs for biological applications should have the following characteristics. First, the coating molecules should be biocompatible and non-toxic. Second, the coating should be stable enough without detachment from nanotube surface in biological solutions especially in serum that have high salt and protein contents. The amphiphilic coating molecules should have very low CMC values so that the nanotube coating is stable after removal of most excess coating molecules from the CNT suspension. Lastly, the coating molecules should have functional groups which are available for bioconjugation with antibodies or other molecules to create various functional CNT conjugates for different biological applications.

Non-covalent functionalization of SWNTs by PEGylated phospholipids (PL-PEG) was developed by our group to meet the requirements above, including high water solubility of nanotubes and versatile functionalities (Fig. 3c) [18, 20, 35, 37, 44]. Phospholipids are the major component of cell membranes, and are safe to use in biological systems. The two hydrocarbon chains of the lipid strongly anchor onto the nanotube surface with the hydrophilic PEG chain extending to the aqueous phase, imparting water solubility and biocompatibility. Unlike nanotubes suspended by typical surfactants, PEGylated SWNTs prepared by this method are highly stable in various biological solutions including serum, and even under harsh conditions without the presence of excess PL-PEG (e.g. stable without coating detachment upon heating in phosphate buffered saline at 70$^{o}$C for weeks). PL-PEGs with different PEG lengths and structures (linear vs branched) can be used to obtain various PEGylated SWNTs for desired applications. Conjugation of biological molecules can be done by using the functional group (e.g. amine) at the PEG terminal. Relying on this functionalization strategy, we have succeeded in using SWNTs for a range of biomedical applications including biological sensing, imaging and drug delivery *in vitro* with cells or *in vivo* with animals [18, 20, 24, 35, 37, 44, 45, 51].

# 3. Selective protein-protein interactions on CNTs and biosensor applications

Functionalization strategies as presented above suggest that both single-walled and multi-walled carbon nanotubes may present scaffolds for biomolecule immobilization, allowing subsequent applications in biosensing, utilizing the intrinsic electronic or optical properties of CNTs for signal transduction. The remarkable physical properties of carbon nanotubes, including high surface area, semiconducting behavior, band-gap fluorescence, and strong Raman scattering spectra, lend themselves well to measuring or detecting proximal or adsorbed biomolecule interactions along the carbon nanotube sidewall, at functionalized cap regions [80], and even within the nanotube shell [81]. Proximity of reasonably charged or polarized biomolecules yields gating effects on isolated semiconducting carbon nanotubes, or net semiconducting networks of CNTs, thus yielding field-effect transistors (FETs) capable of quantifying the degree of specific or non-specific binding of biomolecules [14]. Moreover, the photoluminescent and Raman scattering properties of single-walled carbon nanotubes may be applied to biosensing, by specific conjugation of targeting ligands to SWNT tags, coupled with sufficient sidewall passivation in order to prevent non-specific binding (NSB). Owing to their length scale and unique structure, carbon nanotubes are of great interest in developing highly sensitive and multiplexed biosensors for applications from the laboratory to the clinic.

## 3.1. Non-specific and specific protein-nanotube interactions

In 1999 Balavoine et al reported the crystallization of streptavidin, a biotin-binding protein expressed in *Streptomyces avidinii*, on the hydrophobic surface of MWNTs. While stochastic binding of the ~ 70 kDa protein to the nanotube sidewall was the major observation, helical packing of a streptavidin monolayer was observed on occasion [82]. A similar observation was found for the HupR protein, derived from *Rhodobacter capsulatis*, suggesting that non-specific interactions between proteins and carbon nanotubes are common for such water-soluble biomolecules.

Our group was the first to report that the interaction between proteins and the CNT sidewall is general, and may be ascribed to hydrophobic interactions between the exterior fullerene surface and regions of high hydrophobic residue density within the protein tertiary structure. As such, non-specific adsorption of proteins does not appear to be dependent upon protein pI and occurs in a variety of buffers and solvents [14, 70, 83, 84]. To date, non-specific adsorption of a variety of proteins, ranging in size and pI, has been observed on both single and multi-walled carbon nanotubes (Fig. 4).

The mechanism by which water-soluble proteins interact with bare carbon nanotubes was elucidated, both by the discovery of peptides binding directly to fullerene surfaces [85, 86], and by discovering methodologies to prevent protein adsorption. Our group demonstrated that amphiphilic coatings, containing poly(ethylene glycol), such as Tween-20 and pluronics, were able to effectively decrease the hydrophobicity of the CNT sidewall, thus reducing and often eliminating, the non-specific adsorption of proteins [32, 83, 87]. This result suggested that non-specific protein-CNT interactions were highly dependent upon hydrophobic interactions [88], which were thermodynamically unstable in the presence of surfactants.

A variety of polymers and amphiphiles have been demonstrated for CNT passivation, including non-ionic molecules such as PL-PEGs [32], Triton X-100 [83], Tween-20, and pluronics [14], all of which contain PEG components. Resistance to protein adsorption appears to scale with both the mass of the hydrophilic PEG block, as well as the number of PEG branches. Surfactant-coated carbon nanotubes immobilized on substrates demonstrate degrees of non-specific protein adsorption, ranging from partial inhibition of NSB [14] imparted by Triton X-100 (containing linear PEG $M_n \sim 400$ Da) to full inhibition by Tween-20 (containing branched PEG, $M_n \sim 900$ Da) [14, 83]. Carbon nanotubes that are individualized and suspended in the aqueous phase present significantly higher sterically available surface area (by mass) for non-specific protein interactions, and thus require greater amounts of PEGylation for sufficient passivation. It was experimentally observed [32] that PL-PEG amphiphiles with PEG $M_n \sim 2000$ Da did not efficiently prevent non-specific protein adsorption, however longer linear PEG ($M_n \sim 5000$ Da) or branched PEG ($M_n \sim 8000$ Da) prevented most interactions.

While carbon nanotube-protein interactions can be observed via electron and force microscopies, utilization of such supramolecular systems for biosensing requires that protein activity and specificity remain intact. A comparative study between the enzymes soybean peroxidase and α-chymotrypsin non-specifically adsorbed onto SWNTs revealed varying degrees of specific activity loss, seemingly proportional with protein melting temperature and loading density [89]. Loss of specific enzyme activity was correlated with inverse proportional changes of α-helices (increased) and β-sheets (decreased) composition upon adsorbing onto hydrophobic SWNTs.

For applications of molecular recognition and readout based upon carbon nanotubes, ligand-target affinity must be maintained. Thus, the non-specific immobilization of analyte proteins onto the hydrophobic nanotube sidewall is not desirable, owing to loss of function and lack of chemical control. As described above, several methods for specific nanotube functionalization exist, however covalent modifications to the nanotube sidewall damage the interesting electronic and optical properties expected to be utilized as readout mechanisms in biomolecular detection.

In order to impart chemical functionality to SWNTs immobilized on substrates or suspended in aqueous media (a requirement for most biosensing applications) without hindering electrical or optical readout, non-covalent approaches are critical. Amphiphiles have been employed to impart robust functionality to NTs as noted above, and ideal candidates for biosensing applications must simultaneously repel non-specific interactions and biofouling by including an inert hydrophilic spacer such as poly(ethylene glycol).

Specific protein-nanotube conjugates, synthesized via supramolecular chemistry, have been reported in both surface-immobilized and solution phase methodologies. A proof-of-principle study demonstrated the chemically directed conjugation of streptavidin to bare SWNTs on Si/SiO$_2$, while non-specific adsorption was prevented by a combination of Triton X100 and PEG [83]. Additionally, the surfactant tween-20, adsorbed onto substrate-bound nanotubes, was covalently conjugated via carbonyldiimidazole to biotin, *S*taphylococcal protein A (SpA), and U1A antigen [14, 70], which imparted specific binding of streptavidin, immunoglobulin G, and the monoclonal mouse antibody 10E6 respectively, while preventing non-specific adsorption of non-target proteins.

## 3.2. Electrical detection of protein-nanotube interactions: applications and mechanism

Following the first demonstration of SWNTs acting as molecular sensors via field-effect doping by the gaseous species $NH_3$ and $NO_2$ [90], our lab amongst other set out to utilize field-effect transistors based upon carbon nanotubes for label-free, highly specific biomolecule detection. Conjugation of biotin, *S*taphylococcal protein A (SpA), and U1A antigen to SWNTs [14, 70], amongst other ligands with specific binding characteristics, has been reported to impart specific binding of streptavidin, immunoglobulin G, and the monoclonal mouse antibody 10E6 respectively, allowing *in situ* direct detection of these analytes in the nM range via NT FET devices (Fig. 5).

Surprisingly, in all cases of CNT FET-based detection, conductance across a seminconducting network of NTs was reduced following specific analyte protein binding, regardless of analyte pI [14, 87]. Further studies revealed the importance of Schottky barrier modulation as a result of biomolecule adsorption/binding at nanotube-metal electrode contacts in detecting biomolecules with low net charge [87], thus elucidating the unexpected gating behaviors observed in FET biomolecule detection.

Understanding the Schottky barrier modulation mechanism of carbon nanotube FETs in biomolecule detection allowed for further improvements to be made in device architecture, thus allowing greater sensitivity and lower limits of detection to be coupled with specific binding, imparted by supramolecular ligand conjugations. Angular shadow mask deposition of Au/Cr metal contacts onto semiconducting SWNT networks increased the contact area and reduced the contact thickness, in hopes of increasing the Schottky barrier modulation component in FET conductance measurement of biomolecules (Fig. 6). As hypothesized, increased Schottky barrier area was correlated with improved sensor device characteristics, including a reduction in specific detection limit by four orders of magnitude, over a ~1 μM to 1 pM range of analyte protein concentration [91].

A variety of carbon nanotube FETs and electrochemical devices have been demonstrated for selective detection of oxidase and dehydrogenase activity, as well as for other enzymes and biomolecules of interest, in label-free fashion [92, 93]. However the sensitivity of such devices in performing detection of biomolecules is limited by charge screening of both CNT-bound ligands and ions common in physiological buffers.. Schottky barrier modulation and chemical gating effects of FETs are distant dependent processes, thus reducing the utility of CNT FETs for complex (indirect) bioassays. Especially troublesome for FET applications in detection of biomolecules is the incompatibility of common aqueous buffers with conductance modulated readout. The high ionic strength of physiological buffers, often required to retain protein structure and function, shields FET devices from charge-effects of analyte biomolecules, thus reducing the simplicity and utility of this detection strategy. As a result, detection of analyte concentrations in the nM to pM range has yielded poor signal-to-noise ratios with these devices.

**3.3 Photoluminescent detection of proteins based upon seminconducting SWNTs**

Exploiting the interesting optical properties of semiconducting SWNTs, rather than monitoring their conductance in transistor devices, opens another route for sensitive and selective biomolecule detection using these extraordinary materials. SWNT band gap fluorescence [23] has been explored as a methodology for near infrared-imaging of both *in vitro* and *in vivo* biological systems [17, 24, 26, 50, 94, 95], and holds promise for *in vitro* biomolecule detection assays in both direct and sandwich-assay formats [96-98]. SWNT NIR fluorescence does not photobleach under high excitation powers and benefits from the negligible auto-fluorescence contributions of other assay components in the NIR range. SWNT fluorescence also demonstrates a large Stokes' shift compared with traditional fluorophores, and allows a range of excitation energies to be used.

As described above, aqueous phase processing of SWNTs is made possible by PEGylated amphiphiles, which reduce biofouling and provide sites of functionality for ligand conjugation to the SWNT tags. Recently, our group has demonstrated specific and sensitive detection of biomolecules in sandwich-assay format, using the band-gap photoluminescence of SWNTs as NIR fluorophores, with sensitivity from the micromolar to the picomolar range (unpublished result). SWNT NIR fluorescence may be multiplexed via microarray printing technology, and unique SWNT fluorophores may be obtained by chirality separation [74]. NIR fluorescence detection via SWNT fluorophores may facilitate high throughput bioassays with low background contributions and thus improved sensitivity over conventional techniques.

In addition to direct measurement of SWNT fluorophore emission in immunoassay formats, Strano and coworkers have sought to use band gap modulation and charge transfer effects via photoluminescence for transduction and quantification of biomolecules such as DNA and glucose. Non-covalent functionalization of SWNTs with 24-mer ssDNA and subsequent hybridization of cDNA in proximity to the SWNT surface alters the dielectric constant at the SWNT surface, and produces a 2 meV increase in band gap energy [97], observed as a blue shift in emission. Such methodology yields a theoretical detection limit of 6 nM of 24-mer DNA. Moreover, utilizing a similar non-covalent modification strategy, the same group demonstrated signal transduction via fluorescence-quenching for measuring glucose concentrations at physiologically relevant conditions, from the micromolar to millimolar range [98]. Direct detection of protein binding events by relief of band-gap fluorescence quenching has also been demonstrated. Satishkumar et al demonstrated that small molecule quenchers may be removed from SWNT surfaces by avidin and albumin in a specific and non-specific manner respectively, with detection limits in the micromolar range. [99]

In addition to quenching of the inherent SWNT band-gap fluorescence for sensor applications, specific detection of biomolecules by quenching may also be realized by applying the carbon nanotube as both a scaffold for recognition ligands and as a quencher of small-molecule fluorophores. An excellent example of such quenching-assays is the use of molecular beacons in real-time PCR applications. In comparison with commonly used molecular beacons, SWNTs, non-covalently functionalized by FAM-labeled ssDNA, demonstrate superior quenching in the unhybridzed state, and thus improved single-to-noise ratios when hybridized to a complimentary strand.[100] Additionally, this new class of beacon affords improved thermal stability and is general to a wide variety of fluorophores.

While SWNT NIR band-gap photoluminescence is promising for biomolecule detection, several problems remain to be solved.  While only semiconducting SWNTs demonstrate band gap PL, separation and isolation of these fluorophores from other non-fluorescing SWNT isomers is challenging.  Moreover, the quantum yield of those SWNTs that do fluoresce in the NIR is dependent upon their chemical environment, and processing is required to avoid quenching and maximize quantum yield [101]. Signal transduction via band gap modulation and quenching suffers from the limits of spectral resolution, as well as photoluminescence intensity, which limits the utility of these methods to analytes at relatively high concentration.  Applications of SWNTs as quenchers is promising for detection of ssDNA hybridization, however additional work is necessary to monitor protein interactions by this method.  Future elucidation and optimization of SWNT fluorescence in the NIR range, as well as isolation of semiconducting SWNTs [74, 102] may improve the photoluminescent detection limit of SWNT fluorophores in protein assays by many orders of magnitude.

**3.4 Surface-enhanced SWNT-Raman tags for highly sensitive detection of proteins**

To avoid the issues plaguing SWNT photoluminescence-based detection of proteins in vitro, our lab sought to utilize the intense Raman scattering cross-section of SWNTs in immunoassay format.  While methodologically similar to fluorescence-based sandwich assays [103-105] the application of SWNT Raman tags in lieu of traditional or non-traditional [106] fluorophores holds many potential benefits.  In addition to scattering efficiencies that rival the quantum yield of organic fluorophores [26] the Raman scattering spectra of SWNTs are simple, with strong, well-definied Lorenztian peaks of interest, demonstrating FWHM of 1-3 nm. As such, the Raman scattering spectra of SWNTs are easily distinguishable from noise, and no "auto-scattering" is observed for conventional assay surfaces or reagents [32]. Photobleaching and of SWNT Raman tags is not observed even under extraordinarily high laser powers, a reflection of the stability of the SWNT $sp^2$ carbon lattice.

Surface-enhanced Raman scattering (SERS) [106] is a technique that may be applied to vastly increase the intensity of Raman active molecules in proximity to appropriately tuned surface plasmons, usually associated with gold, silver, or copper nanostructures [107]. Coupling the intense resonance-enhancement of 1D SWNT Raman tags with SERS presents the opportunity to extend the limit of detection of traditional fluorescence assays from approximately 1 pM [108] to the femtomolar level or below. By fabricating a gold-coated assay substrate via electron-beam evaporation and roughening the gold surface via annealing at 400 ºC in $H_2$ following analyte and tag binding, our group was able to demonstrate quantitative SERS over a large area. While such treatment would destroy most small, organic Raman active molecules, SWNTs are robust and are undamaged by the process. This strategy for SERS yielded nearly a 100-fold increase in SWNT Raman scattering intensity [32](Fig. 7).

As previously discussed, non-covalent surfactant wrapping may be used to both functionalize and passivate SWNTs. Recently our group employed SWNTs, suspended by linear PL-PEG-NH$_2$ (M$_n$ ~ 5000 Da) and branched PL-PEG-NH$_2$ (M$_n$ ~ 8000 Da) (Fig. 7), coupled to goat anti-mouse immunoglobulin G (GaM-IgG) for specific detection in immunoassays. To test the selective binding and non-specific behavior of such conjugates, a variety of analyte proteins were immobilized on SERS active substrates, and, following incubation with GaM-IgG-SWNTs, selective detection of mouse IgGs was observed (Fig. 8). Extending this methodology to concentration-dependent sandwich immunoassays, we have been able to push limits of detection to 1 fM of model analyte, approximately three orders of magnitude better than common fluorescence methods. Ultra-sensitive protein detection by SWNT Raman tags has been demonstrated not only for model analyte, but for true biomarkers of human autoimmune disease [32] and cancer (our unpublished results), with dynamic ranges over 6-8 orders of magnitude (Fig. 8).

Biomolecule detection by SWNT Raman tags appears to be generalizable to systems other than high affinity antigen-antibody interactions, including biotin-streptavidin binding, Protein A/G-IgG interaction and DNA hybridization [32]. By coupling the characteristic intense Raman scattering efficiency of 1-D SWNTs with quantitative SERS substrates and non-covalent strategies for both specific antibody conjugation and non-specific binding passivation highly sensitive biosensors have been developed and demonstrated. Moreover, by taking advantage of highly multiplexable microarray technology, and employing isotopically-labeled SWNTs (composed of pure $^{12}$C and $^{13}$C respectively) our group has demonstrated multi-color detection of multiple analytes simultaneously, utilizing only a single excitation source (Fig. 9). The application and utility of such sensitive, multiplexed biosensors remains to be fully explored, though with recent advances in biomarker discovery [109] there is much promise for detection and monitoring of early stage cancer.

## 4. Toxicity of carbon nanotubes

Safety is the first requirement of any material used in medicine. A large number of studies have been performed in the past several years to explore the potential toxic effects of carbon nanotubes. The conclusions of these reports varied drastically, showing a large dependence on the type of nanotube materials as well as functionalization approaches. Cell culture experiments and *in vivo* pilot studies conducted by various groups observed no obvious toxicity of properly functionalized carbon nanotubes [15, 53, 58, 72]. On the other hand, raw carbon nanotubes were shown to be toxic to mice after inhalation into the lung [39, 40, 110, 111]. Recent research showed that unfunctionalized, long MWNTs may pose a carcinogenic risk in mice [43]. As a result of the wide variety of reports, both the public and research communities are currently concerned about using carbon nanotubes for biomedical applications. It is thus critical and urgent to clarify the toxicity issue of carbon nanotubes. The current status is that toxicity appears to be dependent on the material preparation, especially geometry and surface functionalization. Well-functionalized CNTs with biocompatible surface coatings have been shown to be non-toxic *in vitro* to cells and *in vivo* in mice.

**4.1 *In vitro* toxicity of carbon nanotubes**

Even in cell culture experiments, the issue of toxicity of carbon nanotubes is still controversial. While inhibition of HEK 293 cell proliferation after exposure to SWNTs was observed by Cui et al. [38], Ding et al. observed that MWNTs induce cell cycle arrest and increase apoptosis/necrosis of human skin fibroblasts [41]. However, neither of these studies used functionalized carbon nanotubes in the experiments. Apoptosis of T lymphocytes induced by oxidized MWNT was observed by Bottini et al. [42]. However simple oxidization is not enough to render carbon nanotubes soluble and stable in saline and cell medium, and thus dose not represent a biocompatible functionalization. Sayes et al. further reported that the toxicity of CNTs was dependent on the density of functionalization on nanotubes, with minimal toxicity for those heavily functionalized with the highest density of phenyl-$SO_3$X groups on nanotubes [112]. These results are understandable because CNTs without proper functionalization have a highly hydrophobic surface, and thus may aggregate in the cell culture and interact with cells by binding to various biological species including proteins via hydrophobic interactions, to induce certain cell responses such as cell toxicity.

Other factors may also contribute to the observed toxicity of CNT samples *in vitro*. Excess surfactants in the CNT suspensions are known to be highly toxic to cells [113]. The metal catalyst content in CNTs should also be considered when the toxicity of carbon nanotubes is investigated [114]. Moreover, proper assays must be employed in toxicity tests to avoid interference of carbon nanotubes with the assay reagents [115, 116]. For these reasons, *in vitro* toxicity assays of carbon nanotubes should be carefully designed and performed with well prepared and characterized materials, as well as suitable assay methods.

We and many other groups successfully used well functionalized, serum stable carbon nanotubes for *in vitro* cellular uptake experiments without observing apparent toxicity [17, 18, 20, 51-57]. Cells exposed to SWNTs, PEGylated by various PL-PEG amphiphiles used in our work, exhibited neither enhanced apoptosis/neurosis, nor reduced proliferation of various cell lines *in vitro* [18, 20, 51]. Carbon nanotubes covalently functionalized by 1,3-dipolar cycloaddition developed by Prato et al. also appeared to be safe to their tested cell lines, including primary immune cells [52, 53]. Carbon nanotubes with biomimic coating engineered by Bertozzi et al. were also non-toxic to cells [54, 55]. Several other independent groups also reported that CNTs coated by DNA, amphiphlic helical peptides and serum proteins were not toxic to cells [26, 56, 57]. The latest finding by Jin et al. discovered that SWNTs taken up by cells via endocytosis exited cells through exocytosis without affecting the viability of cells [95]. It appears that raw CNTs and CNTs without serum-stable functionalization show toxicity to cells at moderate dosage, while serum-stable, functionalized CNTs show little toxicity even at high dosages.

**4.2 *In vivo* toxicity of carbon nanotubes**

To address the possible side effects of CNTs on human health and our environment, researchers have investigated the toxicology of CNTs in animal models. Unfucntionalized raw CNTs have been intratracheally (IT) instilled into animals, showing obvious pulmonary toxicity including unusual inflammation and fibrotic reactions due to the aggregation of hydrophobic raw CNTs in the lung airways [39, 40, 110, 111]. Those results suggest that aerosol exposure of raw CNTs in the workplace should be avoided to protect human health. Nevertheless, toxicities observed by intratracheal instillation of large amounts of raw CNTs may have little relevance to the toxicology profile of functionalized soluble CNTs for biomedical applications, especially when they are administered through other routes such as intraperitoneal (IP) and intravenous (IV) injections, by which lung airways are not accessible to CNTs.

In a recent pilot study, Poland et al. noticed asbestos-like pathogenic behaviors such as mesothelioma associated with exposing the mesothelial lining of the body cavity of mice to large MWNTs (length 10~50 μm, diameter 80~160 nm) following intraperitoneal injection [43]. Despite the importance of this finding in the discovery of potential negative effects of CNTs to human health, the MWNT materials used in this study are simply sonicated in 0.5% bovine serum albumin (BSA) solutions without careful surface functionalization and hence are not directly meaningful for functionalized CNTs with biocompatible coatings recommended for biomedical applications. Furthermore, length-dependent pathogenicity was observed, as no obvious toxic effect was observed for shorter and smaller MWNTs (length 1~20 μm, diameter 10~14 nm), indicating that the toxicology profiles of CNTs may significantly differ between CNTs of various sizes (diameter and length). It is worth note that functionalized SWNTs used in typical biomedical research have length 50~300 nm and diameter 1~2 nm, which are entirely different from the geometry of MWNTs used by Poland et al.

The first reported in vivo toxicity study of functionalized SWNTs was conducted by the Gambhir group and our group [58]. Both covalently and non-covalently PEGylated SWNTs were used in this study. Mice intravenously injected with PEGyalted SWNTs (~3 mg/kg) were monitored over four months, with systolic blood pressure, complete blood counts and serum chemistry recorded every month. Careful necropsy and tissue histology examinations were performed at the end of 4 months. Normal blood chemistries and histological observations were observed in this study, suggesting that functionalized biocompatible SWNTs may be safe for in vivo biological applications. Another separate study by our group showed similar results, suggesting that PEGylated SWNTs are slowly excreted from the body following systemic distribution in mouse models, without exhibiting obvious toxicity in the process [45]. Recently, Yang et al. showed in a 3 month toxicity study that SWNTs suspended by Tween-80, which is likely not an ideal coating molecule, exhibited low toxicities to the tested mice at a very high dose (~40mg/kg) following i.v. administration. Such toxicity may be due to the oxidative stress induced by SWNTs accumulated in liver and lung [117]. The toxicity observed was dose dependent, and appeared to be less obvious at lower doses (2mg/kg and 16mg/kg). Another recent report by the same group showed that their covalently PEGylated SWNTs, with much higher aqueous stabilities and biocompatibilities, exhibited an ultra-long blood circulation half-life in mice [118]. Although the long-term toxicology of such "improved" SWNTs has yet to be determined, no acute toxicity has been reported even at a high dose (24mg/kg).

To fully address the toxicity concern of CNTs, further investigations, including animal models other than mice, and at larger scales, are still required. Moreover, the interactions between administered CNTs and the immune complement system, whose activation is an important first defense line against foreign species, especially microbes, requires more attentions [119]. Moreover, increased efforts are needed not only from the chemical approach, i.e. further optimizing CNT surface chemistry and geometry for improved biocompatibility, but also from those with biological expertise, to systematically study the complete CNT toxicology profile in different animal models with different routes of administration.

## 5. *In vitro* delivery of biomolecules by carbon nanotubes

The work of using carbon nanotubes for drug delivery in our group was triggered by an unexpected finding that functionalized CNTs are able to enter cells by themselves without obvious toxicity [15]. Similar results were published by the Prato group around the same time [36]. The CNT cellular uptake mechanism may differ depending on the functionalizations and sizes of the CNTs, including endocytosis as reported by us and several other groups [15, 26, 34, 56, 95], or passive diffusion as observed by the Prato group when CNTs are functionalized by 1,3-dipolar cycloaddition [16, 120]. CNTs have been used to efficiently shuttle various biological cargoes, ranging from small drug molecules to bio-macromolecules, such as protein and DNA/RNA, into different types of cells. Once taken up by cells via endocytosis, SWNTs are able to exit cells through exocytosis [95].

**5.1 Delivery of small drug molecules by carbon nanotubes**

Small drug molecules can be covalently conjugated to CNTs for in vitro delivery. Fluorescent dyes and drug cargos were simultaneously linked to 1,3-dipolar cycloaddition functionalized CNTs via amide bonds for the delivery of an anti-cancer drug [69] or an anti-fungi drug [52] into cells. In collaboration with our group, Feazell and Lippard used non-covalently PEGylated SWNTs (by PL-PEG, $M_n\sim2000$ Da) as a longboat delivery system to internalize a platinum(IV) complex, a prodrug of the cytotoxic platinum (II), into cancer cells [121]. The inert platinum(IV) prodrug compounds developed by the Lippard group are activated only after being reduced to the active platinum(II) form. SWNTs tethered with the platinum(IV) complexes through peptide linkages are taken into cancer cells by endocytosis and reside in cell endosomes, where reduced pH induces reductive release of the platinum(II) core complex, thus killing the cancer cells. The cytotoxicity of the platinum(IV) complex increases over 100-fold after attachment to SWNTs. We have also conjugated paclitaxel, a commonly used anti-cancer drug, to branched PEG-coated SWNTs via a cleavable ester bond [37]. The SWNT-PTX conjugate was tested both *in vitro* and *in vivo*.

Beside covalent conjugation, a novel non-covalent supramolecular chemistry has been uncovered in our lab, for loading aromatic drug molecules to functionalized SWNTs by π-π stacking (Fig. 10a&b) [18]. Doxorubicin, a commonly used cancer chemotherapy drug, can be loaded on the surface of PEGylated SWNTs with remarkably high loading, up to 4 gram of drug per 1 gram nanotube, owing to the ultra-high surface area of SWNTs. The loading/binding is pH dependent and favorable for drug release in endosomes and lysosomes, as well as in tumor micro-environments with acidic pH (Fig. 10c). Similar drug loading behaviors have been reported for MWNTs [122], single-walled carbon nanohorns [123] and nano-graphene oxide [124, 125]. The supramolecular approach of drug loading on CNTs opens new opportunities for drug delivery.

Targeting ligands including folic acid [20, 126], peptides [18, 44] and antibodies [24, 127-129] have been used to target CNTs to specific types of cells *in vitro* or to tumors *in vivo*. Targeted drug delivery with CNTs requires conjugation of both targeting molecules and drug molecules to the same nanotube, and thus requires carefully designed strategies [18, 126]. In the work reported by Dhar et al., folic acid (FA) was linked to a Pt(IV) prodrug compound, and then conjugated to PEGylated SWNTs [126], yielding a SWNT-Pt(IV)-FA conjugate that showed enhanced toxicity to folate receptor (FR) positive cells but not to FR negative cells as the result of FA targeted delivery. For the delivery of aromatic drugs such as doxorubicin, which are directly loaded on the nanotube surface via π-π stacking, the functional groups on the SWNT coating molecules (e.g. PL-PEG-amine) can be conjugated with targeting molecules such as Arg–Gly–Asp (RGD) peptide for targeted delivery (Fig. 10a) [18].

Besides drug conjugation and loading outside nanotubes, the hallow structure of CNTs may allow the encapsulation of drug molecules inside nanotubes for drug delivery. Fullerene balls[130], metal ions[131], small compounds such as metallocenes [132], and even DNA molecules[133] have been encapsulated inside CNTs. Although a number of theoretical modeling studies predicted the insertion of biomolecules including chemotherapy drugs[134, 135] into CNTs, drug delivery by encapsulation of drugs inside CNTs has been rarely reported. Further experimental explorations are still needed to examine the possibility of the encapsulation strategy in CNT based drug delivery.

**5.2 Delivery of biomacromolecules by carbon nanotubes**

Unlike various small drug molecules which are able to diffuse into cells, biomacromolecules including proteins, DNA and RNA rarely cross cell membranes by themselves. Intracellular delivery is thus required in order to use those molecules for therapeutic applications. Proteins can be either conjugated or non-covalently absorbed on nanotubes for intracellular delivery [15, 33]. The hydrophobic surface of partially functionalized SWNTs (e.g. oxidized SWNTs) allows non-specific binding of protein. After being translocated into cells by nanotubes, proteins can become bioactive once they are released from endosomes [33].

CNTs can be modified with positive charges to bind DNA plasmids for gene transfection [27-30]. Pantarotto et al. and Singh et al. used amine-terminated SWNTs and MWNTs functionalized by 1,3-dipolar cycloaddition to bind DNA plasmids, and have achieved reasonable transfection efficiency [27, 29]. In the work of Gao et al., amine groups were introduced to oxidized MWNTs for DNA binding and transfection, successfully expressing green fluorescence protein (GFP)-expressing mammalian cells. Although the MWNT based method was less efficient than commercial gene transfection agents, such as lipofectamine 2000, the MWNTs exhibited much lower toxicity [30]. In another study carried out by Liu et al., polyethylenimine (PEI) grafted MWNTs were used for DNA attachment and delivery, which afforded comparable efficacy to the standard PEI transfection method with the benefit of reduced cytotoxicity [28].

Small interfering RNA (siRNA) is able to silence specific gene expression via RNA interference (RNAi) and has generated a great deal of interest in both basic and applied biology [136]. Although viral based siRNA delivery method has shown promise in animal models as well as clinical trials, the safety concern of viral vectors is signficant. It is thus important to develop non-viral vectors for siRNA delivery [137, 138]. With a cleavable disulfide bond linkage between siRNA and SWNTs, we successfully delivered siRNA into cells by nanotubes and observed gene silencing effect (Fig. 11a) [35]. We further showed that our SWNT based siRNA delivery was applicable to those hard-to-transfect human T cells and primary cells, which were resistant to delivery by conventional cationic liposome-based transfection agents (Fig. 11 b&c).[51] Surface functionalization dependent cell uptake of SWNTs was observed. Compared with SWNTs coated with long PEG (5.4 kDa), shorter PEG (2 kDa) coated SWNTs with more hydrophobic surface exposed showed higher cellular uptake, which was favorable for siRNA delivery into cells (Fig. 11d). We proposed that our SWNTs functionalized with short PEG (2kDa) retained certain hydrophobicity (due to incomplete coverage of nanotube sidewalls), which could cause binding and association with cells, resulting from hydrophobic interactions with hydrophobic cell membrane domains. The cell binding of SWNTs is an important first step for cellular entry via endocytosis. Our results suggest that balanced chemical functionalization schemes that impart sufficient aqueous solubility and biocompatibility to nanotubes, and retain the ability of nanotube binding with cell surfaces are important for intracellular delivery of biomacromolecules by CNTs. Beside our work, other CNT based siRNA delivery has also been reported, showing efficacy in vitro and even in vivo [139].

## 6. *In vivo* biodistribution and long term fate of carbon nanotubes

Encouraged by the successes of using CNTs for *in vitro* sensing, drug delivery and imaging, research in this field has moved to *in vivo* animal research. The first critical question to address is the biodistribution profile of CNTs after systemic administration into animals and any toxicity. In the past few years, *in vivo* biodistribution and pharmacokinetics studies were carried out by a number of groups using different CNT materials, different surface functionalizations, and different tracking methodologies, thus obtaining various and sometimes controversial results.

Radio-labeled ($^{111}$In-DTPA) SWNTs and MWNTs functionalized by 1,3-dipolar cycloaddition were used by Singh et al.[46] and Lacerda et al.[47] to determine biodistribution. Surprisingly, after intravenous injection of CNTs into mice, they observed fast urinal clearance of CNTs, with the majority (>95%) cleared out within 3 hours, and no uptake in RES organs such as the liver and spleen. Those phenomena were similar to the in vivo behaviors of small molecules, but drastically differed from that expected of most nanoparticles with sizes exceeding the glomerular filtration threshold. To explain their results, the researchers proposed that despite finite lengths, the small diameters of CNTs allowed for fast urinal excretion of CNTs. However, this conclusion is debatable considering well-characterized size dependent protein biodistribution and excretion behaviors (Table 1, proteins larger than 7-9nm start showing high RES uptake and limited renal excretion) and also contradictory findings in a study reported by Choi et al. using quantum dots (QDs) [140]. It is found that the size limitation of spherical QDs to undergo fast urinal excretion is ~6 nm, including coating molecules [140], which is indeed larger than the diameter of individual SWNTs (1-2nm). However, the QDs were but much smaller than the diameter of SWNT bundles (10-40nm) [46] or MWNTs (20-30nm) [47] used in these two CNT biodistribution studies. The reported fast CNT urinal excretion requires confirmation.

Several other labs have also studied the biodistribution of radio-labeled CNTs in mice. Wang & Liu et al. reported relatively slow urinal excretion and low RES uptake in their first study [141]. Later reports by the same group using $^{14}$C-taurine functionalized CNTs, however, revealed dominant and persistent liver accumulation of CNTs after intravenous injection [49, 142]. Another independent study by McDevitt et al. using antibody conjugated radio-labeled CNTs functionalized by 1,3-dipolar cycloaddition also showed high CNT uptake in the liver and spleen with slow urinal excretion [129]. A significant amount of CNTs remained in the body even after 15 days. We have also investigated the biodistribution of radio-labeled, PEGylated SWNTs, observing dominant SWNT uptake in RES organs, including liver and spleen, without rapid clearance [44]. Although the radiolabel method is a convenient way to examine the biodistribution of a substance, excess free radio isotopes in the radio-labeled CNT samples, if not completely removed, may lead to false results, especially for excretion as free radio isotopes are small molecules that are rapidly excreted through urine after intravenous injection. Also, radiolabels could gradually release from CNTs *in vivo,* and be slowly excreted in the free form. As a consequence, radio-labeling is not an ideal strategy to study the excretion and long term fate of CNTs.

Without using radiolabels, researchers have investigated the in vivo behaviors of SWNTs relying on their intrinsic properties. Individual semi-conducting SWNTs exhibit

NIR photoluminescence, which has been used by Cherukuri et al. to track nanotubes in rabbits [50]. Without detailed biodistribution data, the authors observed SWNT photoluminance signals in the liver but not in other organs such as kidney. In another study, Yang et al. used isotope ratio mass spectroscopy to examine the biodistribution of $^{13}$C enriched unfunctionalized SWNTs over a month, showing high nanotube uptake in lung, liver and spleen without apparent excretion within 28 days [48]. Taking advantage of the bright Raman scattering signatures of SWNTs, we used Raman spectroscopy to study the long-term fate of nanotubes in mice [45]. It was uncovered that our PEGylated biocompatible SWNTs were dominantly accumulated in liver and spleen after intravenous administration, but slowly excreted within months, likely via the biliary pathway into the feces. There could be a small portion of SWNTs with very short lengths that were extreted through urinal elimination, as indicated by the weak SWNT Raman signal observed in the mouse kidney and bladder.

Most importantly, we have systematically studied the relationship between PEG coating and the *in vivo* behaviors of SWNTs. Prolonged blood circulation time is generally desired for drug delivery and tumor targeting. We have concluded that long PEG coatings on SWNTs generally confer a prolonged blood circulation half-life, reduced RES uptake and accelerated excretion (Fig. 12). Moreover, PEG with branched structure offers a more efficient coating on SWNT surface than linear PEGs, and allows a longer SWNT blood circulation half-life, which is ~5 h for branched PEG compared with ~2 h for linear PEG at the same molecular weight (7kDa) (Fig. 12) [45]. Another recent study by Yang et al. obtained an even longer blood circulation half-life (22 h) using covalently PEGylated SWNTs [118]. Those results are thus illuminative to the future research of carbon nanotubes for in vivo biomedical applications.

| Protein Molecule | MW (kDa) | d (nm) | Biodistribution (%ID/g at 4 h) | | | Blood half-life (min) | Whole body half-life (h) |
|---|---|---|---|---|---|---|---|
| | | | liver | spleen | kidney | | |
| ScFv [143] | 30 | 5.3 [140] | 1.3 | 0.9 | 0.8 | <10 | ~3.8 |
| Fab' [143] | 50 | 6.0 [144] | 1.3 | 1.8 | 21.5 | ~30 | ~6.7 |
| Sc(Fv)$_2$ [143, 145] | 60 | 7.0 [140] | 2.0 | 2.8 | 2.9 | 78 | ~10 |
| F(ab')$_2$ [143] | 100 | - | 7.8 | 7.1 | 9.8 | ~200 | ~20 |
| [sc(Fv)$_2$]$_2$ [145] | 120 | 9.3 [140] | 7.0 | 6.2 | 3.6 | 170 | - |
| IgG [143, 145] | 152 | 11 [144] | 18.7 | 18.0 | 4.7 | 330 | ~80 |

**Table 1.** Biodistribution and pharmacokinetics of proteins with different sizes. Small proteins with diameter below 6~7 nm are quickly excreted through urinal clearance with very short blood circulation and whole body retention half-lives. Bigger proteins in contrast show high uptake in RES organs and much longer blood and whole body half-lives. They are unable to undergo fast urinal excretion because of their large sizes.

# 7. *In vivo* tumor targeting and preliminary effort towards in vivo cancer therapy

In order to use CNTs for potential cancer treatment and/or imaging, targeting nanotubes to tumors is highly desired. Both passive targeting, relying on the enhanced permeability and retention (EPR) effect of cancerous tumors, and active targeting guided by tumor targeting ligands, have been applied for various nanoparticle-based drug delivery systems. Thus far there are two published papers reporting *in vivo* tumor targeting by CNTs conjugated with targeting ligands. We showed that efficient tumor targeting was achieved by conjugating an Arg–Gly–Asp (RGD) peptide, which recognizes integrin $\alpha_v\beta_3$, up-regulated on various solid tumor cells and tumor vasculatures, to PEGylated SWNTs [44]. SWNTs with two different PEG coatings conjugated with both RGD peptide and radiolabels ($^{64}$Cu-DOTA) were intravenously injected in to glioblastoma U87MG tumor bearing mice, which were monitored by micro-positron emission tomography (micro-PET) over time (Fig. 13). RGD conjugated SWNTs with long PEG coating (SWNT-PEG$_{5400}$-RGD) exhibited a high tumor uptake of ~13 % of injected dose per gram tissue (%ID/g), compared with 4~5 %ID/g obtained with plain SWNTs without RGD (SWNT-PEG$_{5400}$). Interestingly, we uncovered that efficient tumor targeting could only be realized when SWNTs were coated with long PEG (SWNT-PEG$_{5400}$-RGD) but not with short PEG (SWNT-PEG$_{2000}$-RGD). The latter had short blood circulation time, and thus lower probability to be trapped in tumors or bind the tumor receptors. Our data suggest that surface functionalization of SWNTs is also important for tumor targeting *in vivo*. Another study carried out by McDevitt et al.[129] showed tumor targeting of CNTs by antibody conjugation.

The first *in vivo* cancer treatment study with CNTs was reported by Zhang et al. using positively charged SWNTs to delivery therapeutic siRNA into cancer cells [139]. However, this was a proof-of-concept study, with SWNT-siRNA complexes directly injected into tumors, instead of systemic administration. Our recent work showed that paclitaxel (PTX), a commonly used chemotherapy drug, may be conjugated to branched PEG functionalized SWNTs via a cleavable ester bond (Fig. 14a) [37]. The SWNT-PTX conjugate was tested in a 4T1 murine breast cancer model in mice, exhibiting improved treatment efficacy over the clinical cremophor-based PTX formulation, Taxol® (Fig. 14b). Pharmacokinetics and biodistribution studies revealed longer blood circulation half-life and higher tumor uptake of SWNT-PTX than those of simple PEGylated PTX and Taxol®, consistent with the observed efficacies of different PTX formulations. The high tumor passive uptake of SWNT-PTX was likely due to the EPR effect. In addition, PTX molecules carried to RES organs (e.g. liver and spleen) by SWNTs were rapidly dissociated from nanotubes and excreted, diminishing the RES toxicity of this SWNT-based PTX formulation. Our work is the first one to show that carbon nanotubes can be used for *in vivo* drug deliver for cancer therapy by systemic administration [37].

# 8. Biological imaging using carbon nanotubes

In addition to applications for drug delivery and treatment, the intrinsic optical properties of SWNTs make them useful as optical probes. Owing to their quasi 1-D nature, SWNTs exhibit strong resonance Raman scattering, high optical absorption and photoluminescence in the near-infrared (NIR) range, all of which have been utilized for imaging in biological systems in vitro and in vivo.

## 8.1 *In vitro* photoluminescence imaging

Individual semiconducting SWNTs have small bandgaps on the order of ~1eV, depending on the diameter and chirality of a given nanotube. This bandgap allows for photoluminescence in the NIR range (900nm-1600nm) which is useful for biological imaging, due to the high optical transparency of biological tissue near 800nm-1000nm and inherently low autofluorescence from tissue in the NIR range [146]. SWNTs have a further advantage due to the large separation between the excitation (550nm-850nm) and emission bands (900nm-1600nm). This spacing further reduces background from autofluorescence and Raman scattering.

O'Connell et al. [23] first demonstrated NIR photoluminescence from micelle encapsulated SWNTs, yielding an estimated quantum efficiency of about $10^{-3}$. Relying on this intrinsic NIR photoluminescence, Cherukuri et al. imaged nonspecific uptake of SWNTs in phagocytic cells [17]. In more recent work, Jin et al. user NIR photoluminescence to track endocytosis and exocytosis of SWNTs in NIH-3T3 cells in real time [95]. In our lab, we developed bio-inert PEGylated SWNTs conjugated with antibodies as NIR fluorescent tags for selective probing of cell surface receptors [24]. Figure 15 shows NIR fluorescence images of (b) BT-474, which is HER2/neu positive, and (c) MCF-7, which is HER2/neu negative, treated with a SWNT-Herceptin conjugate. We observed ultra-low NIR autofluorescence between different cell lines, demonstrating the advantage of SWNTs as a NIR fluorophore.

There is still work to be done in the application of SWNTs as effective fluorophores. Bio-inert samples must be made with optimized quantum yields. Initial estimates of the quantum yield of SWNT suspensions were on the order of $10^{-3}$ or less [23, 147]. This seemed to contradict single tube studies with much higher quantum yield estimates of up to 0.07 [148]. Many factors contribute to the quantum yield including exciton quenching by bundles [149], sidewall defects [150], and length [151]. Crochet et al. showed that the quantum yield of bulk SWNT suspensions could be increased by an order of magnitude by centrifugation in an iodixanol gradient to remove bundled nanotubes [152]. Careful preparation techniques such as these need to be applied to making suspensions of pristine, unbundled SWNTs in bio-inert coatings, such as PL-PEG, in order to fully realize their potential as sensitive NIR fluorophores.

**8.2 *In vitro* Raman imaging**

Owing to their quasi 1-D nature, SWNTs exhibit strong resonance Raman scattering due to their sharp electronic density of states at the van Hove singularities. SWNTs have several unique Raman scattering features including the radial breathing mode (RBM) and tangential mode (G-band) [25], which are sharp and strong peaks that can be easily distinguished from fluorescence backgrounds, and thus are suitable for optical imaging.

We and Heller et al. have used Raman microscopy to image SWNTs in liver cells, as well as tissue slices, using either RBM peak or G-band peak of SWNTs [26, 45, 51, 58]. Our latest work showed that SWNTs with different isotope compositions exhibited shifted G band peaks, and thus could be used as multi-color contrast agents for multiplexed Raman imaging [127]. Cancer cells with different receptor expression profiles were selectively labeled with three isotopically unique formulations of "colored" SWNTs, conjugated with various targeting ligands including Herceptin (anti-Her2), Erbitux (anti-Her1) and RGD peptide, allowing for multi-color confocal Raman imaging of cells in a multiplexed manner using a single excitation (Fig. 16 c&d). SWNT Raman signals are highly robust against photo-bleaching, allowing long term imaging and tracking [26, 45]. With narrow peak features, SWNT Raman signals are easily differentiated from the auto-fluorescence background. The SWNT Raman excitation and scattering photons are in the near-infrared region, which is the most transparent optical window for biological systems in vitro and in vivo. Thus, SWNTs are novel Raman tags promising for multiplexed biological detection and imaging.

**8.3  SWNTs for *in vivo* animal imaging**

SWNT based biomedical imaging has also been conducted in animal models. The first-ever imaging of nanotubes inside a living animal was achieved by Weisman group in 2007 [94]. In this work, drosophila larvae were fed by food containing SWNTs and imaged by NIR fluorescence microscope. The biodistribution of SWNTs in live larvae was monitored by the nanotube fluorescence signals. Recently, the Gambhir group successfully used RGD-conjugated PEGylated SWNTs provided by us as Raman probes for *in vivo* tumor imaging in live mice (Fig. 17) [153, 154]. Intravenous injection of targeting SWNTs to living mice bearing a tumor xenograft, showed strong SWNT Raman signals in the tumor, while little signal was observed in the tumor upon injection of non-targeted SWNTs. This is the first success of *in vivo* tumor imaging via carbon nanotube labels.

SWNTs have strong optical absorption in the visible and NIR range. We and Chakravarty et al. have shown that SWNTs can be utilized as a photo-thermal therapeutic agents to kill cancer cells [20, 128]. NIR laser irradiation was used in both cases to generate heat, causing destruction of cancer cells with specific SWNT internalization. Beside its potential applications in therapy, the high optical absorption of SWNTs can also be utilized in photoacoustic imaging. Photoacoustic imaging, in which sounds are generated as a result of local heating by the absorption of laser light, has higher spatial resolution than traditional ultra-sound, and deeper tissue penetration than fluorescence imaging [155]. de la Zerda et al. used our RGD conjugated SWNTs as the contrast agent for photoacoustic molecular imaging of cancer in a mouse tumor model (Fig. 17) [22]. This work opens up new opportunities for *in vivo* biological imaging with SWNTs.

## 9. Summary and Prospective

In this article, we have comprehensively reviewed the current research regarding the use of carbon nanotubes for biomedical applications. Various covalent and non-covalent chemistries have been developed to functionalize CNTs for biomedical research. Relying on their electric or optical properties, functionalized CNTs have been used for ultra-sensitive detection of biological species. Surveying the literature, we clarify that *in vitro* and *in vivo* toxicities of CNTs are highly dependent on CNT functionalization. Well functionalized CNTs with biocompatible coatings are stable in biological solutions, and non-toxic *in vitro* to cells and *in vivo* to mice at the tested doses. Various reports have shown that CNTs are able to shuttle biological molecules including small drug molecules and biomacromolecules including proteins, plasmid DNA and siRNA into cells *in vitro* via an endocytosis pathway. *In vivo* behaviors including blood circulation, biodistribution and long term fate of CNTs have been studied in the past two years, showing dominant uptake of CNTs in RES organs, similar to most nanomaterials tested *in vivo*. CNTs are able to target tumors by both passive targeting relying on the EPR effect and active targeting guided by targeting ligands, promising for *in vivo* cancer treatment. Moreover, SWNTs exhibit unique intrinsic optical properties and have been used for biological imaging *in vitro* and *in vivo*.

Among all aspects, surface functionalization chemistry is the most essential and fundamental part for CNT biomedical applications. Highly hydrophilic coating, such as long and branched PEG on CNTs imparts 'inertness' in biological environments, minimizing the toxicity of CNTs and reducing non-specific binding (NSB) to biological species, such as serum proteins and cell surface proteins. Minimal NSB is critical to enhance the detection sensitivity in CNT based biosenors and imaging probes. Decreased NSB reduces the non-specific endocytosis of CNTs, which is favored for targeted drug delivery to specific cell types. Similarly, the *in vivo* behaviors of CNTs are highly dependent on the nanotube surface coating. Prolonged blood circulation and reduced RES uptake can be achieved by using CNTs with highly hydrophilic coatings. Further efforts are required to optimize the surface chemistry of SWNTs, to further enhance biocompatibility. With improved surface coating, further improvements in biological sensing and imaging, better tumor targeting attributed to prolonged blood circulation and reduced RES uptake, and accelerated excretion, may be realized. By minimizing the non-specific protein binding of nanotubes via improved surface functionalization, complement activation may also be reduced for SWNTs .

CNTs, especially SWNTs are highly promising in biomedicine due to several features. CNTs are composed purely of carbon, while many inorganic nanomaterials (e.g. quantum dots) are composed of relatively more hazardous elements, such as heavy metals. The unique 1D structure and tunable length of CNTs provide an ideal platform to investigate size and shape effects *in vivo*. Lastly, unlike conventional organic drug carriers, the intrinsic physical properties of SWNTs including resonance Raman scattering, photoluminescence, and strong NIR optical absorption can provide valuable means of tracking, detecting and imaging. Taken together, CNTs may serve as a unique platform for potential multimodality cancer therapy and imaging.

Although numerous encouraging results of using CNTs in biomedicine have been published in the past several years, much more work is still needed before CNTs can enter the clinic. The most important issue to be addressed is still the concern of long-term toxicity. Although we have shown that well functionalized carbon nanotubes are not toxic *in vitro* to cells and *in vivo* to mice at our tested doses, further systematic investigations using different animal models at larger scales with various doses are required. Special attention should be paid to CNTs with surface functionalization optimized for such applications, with greater chances of minimizing toxic side-effects.

Although we have succeeded in using carbon nanotubes for *in vitro* siRNA delivery, *in vivo* delivery of this type of biological macromolecules remains a challenge. There is a paradox existing due to the need of high *in vitro* cellular uptake and the requirement of favorable *in vivo* behaviors such as long blood circulation and low RES uptake of SWNTs. The former suggests reduced PEGylation of nanotubes, while the latter desires dense biocompatible surface coatings. Conjugation of targeting ligands on well coated SWNTs may help to solve this problem, allowing enhanced cellular uptake via receptor mediated endocytosis, without loss of optimal SWNT *in vivo* characteristics. Further development of suitable bioconjugation chemistry on nanotubes may create versatile SWNT based bioconjugates for actively targeted *in vivo* drug and gene delivery.

Although not specifically discussed in this review, the fabrication of CNTs may also play important roles in their future biomedical applications. Most of CNT samples used in the published biomedical studies are heterogeneous mixtures of nanotubes with different lengths, diameters and chiralities. The lengths of CNTs may affect *in vitro* cellular uptake as well as *in vivo* pharmacokinetics of nanotubes. It is thus important to obtain and test nanotube samples with narrow length distributions. A density column based length separation method was established in our group [156]. Further studies will uncover the potential length dependent effect of CNTs to their behaviors in biological systems. The electronic structures of SWNTs are determined by their chiralities (Fig. 1). Achieving SWNTs with single diameter and chirality may bring a revolution to the semiconductor industry and has been one of the ultimate goals of nanotube research for a decade. Progresses have been made by selective synthesis of SWNTs at special conditions [157], removal of SWNTs with undesired diameters by etching [102], and chirality separation based on chromatography[73, 74]. Different semiconducting SWNTs with single chiralitiy compositions can serve as different colors in the NIR photoluminance imaging (Fig. 1). One the other hand, Raman scattering of SWNTs with single chirality will be largely enhanced because all nanotubes can be in the resonance with a selected excitation wavelength. The diameter dependent Raman RBM band of SWNTs may also be utilized in multi-color Raman imaging [25, 158]. CNT samples with homogenous length, diameter and chirality distributions are the ideal candidates for further studies in biomedicine.

Lastly, biomedical imaging, based on the inherent physical properties of SWNTs, may be combined with drug delivery for multimodality cancer diagnosis and therapy. Phototherapy relying on the strong NIR light absorption ability of SWNTs may be conducted simultaneously, along with chemotherapy delivered by SWNTs, to enhance treatment efficacy *in vivo*. Despite challenges on the way towards the clinic, carbon nanotubes exhibit great potential for biomedicine, and may bring unprecedented opportunities for the future of cancer diagnosis and therapy.

# References

1. Whitesides, G. M. The 'right' size in nanobiotechnology. *Nat. Biotech.* **2003**, *21,* 1161-1165.
2. Lowe, C. R. Nanobiotechnology: the fabrication and applications of chemical and biological nanostructures. *Curr. Opin. Chem. Biol.* **2000**, *10,* 428-434.
3. Wang, L.; Zhao, W. and Tan, W. Bioconjugated silica nanoparticles: development and applications. *Nano Res.* **2008**, *1,* 99-115.
4. Iijima, S. Helical microtubules of graphitic carbon. *Nature* **1991**, *354,* 56-58.
5. Dai, H. Carbon nanotubes: Synthesis, integration, and properties. *Acc. Chem. Res.* **2002**, *35,* 1035-1044.
6. Dresselhaus, M. and Dai, H. (eds.) MRS 2004 Carbon Nanotube Special Issue, Vol. 29. 2004.
7. Golberg, D.; Costa, P. M. F. J.; Mitome, M. and Bando, Y. Nanotubes in a gradient electric field as revealed by STM-TEM technique. *Nano Res.* **2008**, *1,* 166-175.
8. Zhou, W.; Rutherglen, C. and Burke, P. Wafer scale synthesis of dense aligned arrays of single-walled carbon nanotubes. *Nano Res.* **2008**, *1,* 158-165.
9. Ago, H.; Petritsch, K.; Shaffer, M. S. P.; Windle, A. H. and Friend, R. H. Composites of carbon nanotubes and conjugated polymers for photovoltaic devices. *Adv. Mater.* **1999**, *11,* 1281-1285.
10. Javey, A.; Guo, J.; Wang, Q.; Lundstrom, M. and Dai, H. J. Ballistic carbon nanotube field-effect transistors. *Nature* **2003**, *424,* 654-657.
11. Cao, Q. and Rogers, J. A. Random networks and aligned arrays of single-walled carbon nanotubes for electronic device applications. *Nano Res.* **2008**, *1,* 259-272.
12. Fan, S. S.; Chapline, M. G.; Franklin, N. R.; Tombler, T. W.; Cassell, A. M. and Dai, H. J. Self-oriented regular arrays of carbon nanotubes and their field emission properties. *Science* **1999**, *283,* 512-514.
13. Dillon, A. C.; Jones, K. M.; Bekkedahl, T. A.; Kiang, C. H.; Bethune, D. S. and Heben, M. J. Storage of hydrogen in single-walled carbon nanotubes. *Nature* **1997**, *386,* 377-379.
14. Chen, R. J.; Bangsaruntip, S.; Drouvalakis, K. A.; Kam, N. W. S.; Shim, M.; Li, Y. M.; Kim, W.; Utz, P. J. and Dai, H. J. Noncovalent functionalization of carbon nanotubes for highly specific electronic biosensors. *Proc. Nat. Acad. Sci. U. S. A.* **2003**, *100,* 4984-4989.
15. Kam, N. W. S.; Jessop, T. C.; Wender, P. A. and Dai, H. J. Nanotube molecular transporters: Internalization of carbon nanotube-protein conjugates into mammalian cells. *J. Am. Chem. Soc.* **2004**, *126,* 6850-6851.
16. Bianco, A.; Kostarelos, K.; Partidos, C. D. and Prato, M. Biomedical applications of functionalised carbon nanotubes. *Chem. Commun.* **2005**, 571-577.
17. Cherukuri, P.; Bachilo, S. M.; Litovsky, S. H. and Weisman, R. B. Near-infrared fluorescence microscopy of single-walled carbon nanotubes in phagocytic cells. *J. Am. Chem. Soc.* **2004**, *126,* 15638-15639.

18. Liu, Z.; Sun, X.; Nakayama, N. and Dai, H. Supramolecular Chemistry on Water-Soluble Carbon Nanotubes for Drug Loading and Delivery. *ACS Nano* **2007**, *1,* 50-56.
19. Tans, S. J.; Devoret, M. H.; Dai, H. J.; Thess, A.; Smalley, R. E.; Geerligs, L. J. and Dekker, C. Individual single-wall carbon nanotubes as quantum wires. *Nature* **1997**, *386,* 474-477.
20. Kam, N. W. S.; O'Connell, M.; Wisdom, J. A. and Dai, H. Carbon nanotubes as multifunctional biological transporters and near-infrared agents for selective cancer cell destruction. *Proc. Natl. Acad. Sci. USA* **2005**, *102,* 11600-11605.
21. Chakravarty, P.; Marches, R.; Zimmerman, N. S.; Swafford, A. D. E.; Bajaj, P.; Musselman, I. H.; Pantano, P.; Draper, R. K. and Vitetta, E. S. Thermal ablation of tumor cells with anti body-functionalized single-walled carbon nanotubes. *Proc. Natl. Acad. Sci. U. S. A.* **2008**, *105,* 8697-8702.
22. Zerda, A. d. l.; Zavaleta, C.; Keren, S.; Vaithilingam, S.; Bodapati, S.; Liu, Z.; Levi, J.; Ma, T.-J.; Oralkan, O.; Cheng, Z.; et al. Photoacoustic Molecular Imaging in Living Mice Utilizing Targeted Carbon Nanotubes. *Nat. Nanotech.* **2008**, *3,* 557 - 562.
23. O'Connell, M. J.; Bachilo, S. M.; Huffman, C. B.; Moore, V. C.; Strano, M. S.; Haroz, E. H.; Rialon, K. L.; Boul, P. J.; Noon, W. H.; Kittrell, C.; et al. Band gap fluorescence from individual single-walled carbon nanotubes. *Science* **2002**, *297,* 593-596.
24. Welsher, K.; Liu, Z.; D, D. and Dai, H. Selective Probing and Imaging of Cells with Single Walled Carbon Nanotubes as Near-Infrared Fluorescent Molecules. *Nano Lett.* **2008**, *8,* 586-590.
25. Rao, A. M.; Richter, E.; Bandow, S.; Chase, B.; Eklund, P. C.; Williams, K. A.; Fang, S.; Subbaswamy, K. R.; Menon, M.; Thess, A.; et al. Diameter-selective Raman scattering from vibrational modes in carbon nanotubes. *Science* **1997**, *275,* 187-191.
26. Heller, D. A.; Baik, S.; Eurell, T. E. and Strano, M. S. Single-walled carbon nanotube spectroscopy in live cells: Towards long-term labels and optical sensors. *Adv. Mater.* **2005**, *17,* 2793-2799.
27. Pantarotto, D.; Singh, R.; McCarthy, D.; Erhardt, M.; Briand, J. P.; Prato, M.; Kostarelos, K. and Bianco, A. Functionalized carbon nanotubes for plasmid DNA gene delivery. *Angew. Chem. Int. Ed.* **2004**, *43,* 5242-5246.
28. Liu, Y.; Wu, D. C.; Zhang, W. D.; Jiang, X.; He, C. B.; Chung, T. S.; Goh, S. H. and Leong, K. W. Polyethylenimine-grafted multiwalled carbon nanotubes for secure noncovalent immobilization and efficient delivery of DNA. *Angew. Chem. Int. Ed.* **2005**, *44,* 4782-4785.
29. Singh, R.; Pantarotto, D.; McCarthy, D.; Chaloin, O.; Hoebeke, J.; Partidos, C. D.; Briand, J. P.; Prato, M.; Bianco, A. and Kostarelos, K. Binding and condensation of plasmid DNA onto functionalized carbon nanotubes: Toward the construction of nanotube-based gene delivery vectors. *J. Am. Chem. Soc.* **2005**, *127,* 4388-4396.
30. Gao, L. Z.; Nie, L.; Wang, T. H.; Qin, Y. J.; Guo, Z. X.; Yang, D. L. and Yan, X. Y. Carbon nanotube delivery of the GFP gene into mammalian cells. *Chembiochem* **2006**, *7,* 239-242.

31. Tang, X. W.; Bansaruntip, S.; Nakayama, N.; Yenilmez, E.; Chang, Y. L. and Wang, Q. Carbon nanotube DNA sensor and sensing mechanism. *Nano Lett.* **2006**, *6,* 1632-1636.
32. Chen, Z.; Tabakman, S. M.; Goodwin, A. P.; Kattah, M. G.; Daranciang, D.; Wang, X.; Zhang, G.; Li, X.; Liu, Z.; Utz, P. J.; et al. Protein Microarrays with Carbon Nanotubes as Multi-Color Raman Labels. *Nat. Biotech.* **2008**, *in press, DOI: 10.1038/nbt.150*.
33. Kam, N. W. S. and Dai, H. Carbon nanotubes as intracellular protein transporters: Generality and biological functionality. *J. Am. Chem. Soc.* **2005**, *127,* 6021-6026.
34. Kam, N. W. S.; Liu, Z. A. and Dai, H. J. Carbon nanotubes as intracellular transporters for proteins and DNA: An investigation of the uptake mechanism and pathway. *Angew. Chem. Int. Ed.* **2006**, *45,* 577-581.
35. Kam, N. W. S.; Liu, Z. and Dai, H. Functionalization of carbon nanotubes via cleavable disulfide bonds for efficient intracellular delivery of siRNA and potent gene silencing. *J. Am. Chem. Soc.* **2005**, *127,* 12492-12493.
36. Pantarotto, D.; Briand, J. P.; Prato, M. and Bianco, A. Translocation of bioactive peptides across cell membranes by carbon nanotubes. *Chem. Commun.* **2004***,* 16-17.
37. Liu, Z.; Chen, K.; Davis, C.; Sherlock, S.; Cao, Q.; Chen, X. and Dai, H. Drug delivery with carbon nanotubes for in vivo cancer treatment. *Cancer Res.* **2008**, *68,* 6652-6660.
38. Cui, D. X.; Tian, F. R.; Ozkan, C. S.; Wang, M. and Gao, H. J. Effect of single wall carbon nanotubes on human HEK293 cells. *Toxicol. Lett.* **2005**, *155,* 73-85.
39. Lam, C. W.; James, J. T.; McCluskey, R. and Hunter, R. L. Pulmonary toxicity of single-wall carbon nanotubes in mice 7 and 90 days after intratracheal instillation. *Toxicol. Lett.* **2004**, *77,* 126-134.
40. Warheit, D. B.; Laurence, B. R.; Reed, K. L.; Roach, D. H.; Reynolds, G. A. M. and Webb, T. R. Comparative pulmonary toxicity assessment of single-wall carbon nanotubes in rats. *Toxicol. Lett.* **2004**, *77,* 117-125.
41. Ding, L. H.; Stilwell, J.; Zhang, T. T.; Elboudwarej, O.; Jiang, H. J.; Selegue, J. P.; Cooke, P. A.; Gray, J. W. and Chen, F. Q. F. Molecular characterization of the cytotoxic mechanism of multiwall carbon nanotubes and nano-onions on human skin fibroblast. *Nano Lett.* **2005**, *5,* 2448-2464.
42. Bottini, M.; Bruckner, S.; Nika, K.; Bottini, N.; Bellucci, S.; Magrini, A.; Bergamaschi, A. and Mustelin, T. Multi-walled carbon nanotubes induce T lymphocyte apoptosis. *Toxicol. Lett.* **2006**, *160,* 121-126.
43. Poland, C. A.; Duffin, R.; Kinloch, I.; Maynard, A.; Wallace, W. A. H.; Seaton, A.; Stone, V.; Brown, S.; MacNee, W. and Donaldson, K. Carbon nanotubes introduced into the abdominal cavity of mice show asbestos-like pathogenicity in a pilot study. *Nat. Nanotech.* **2008**, *3,* 423 - 428.
44. Liu, Z.; Cai, W. B.; He, L. N.; Nakayama, N.; Chen, K.; Sun, X. M.; Chen, X. Y. and Dai, H. J. In vivo biodistribution and highly efficient tumour targeting of carbon nanotubes in mice. *Nat. Nanotech.* **2007**, *2,* 47-52.
45. Liu, Z.; Davis, C.; Cai, W.; He, L.; Chen, X. and Dai, H. Circulation and long-term fate of functionalized, biocompatible single-walled carbon nanotubes in mice probed by Raman spectroscopy. *Proc. Natl. Acad. Sci. U. S. A.* **2008**, *105,* 1410-1415.
46. Singh, R.; Pantarotto, D.; Lacerda, L.; Pastorin, G.; Klumpp, C.; Prato, M.; Bianco, A. and Kostarelos, K. Tissue biodistribution and blood clearance rates of intravenously administered carbon nanotube radiotracers. *Proc. Nat. Acad. Sci. U. S. A.* **2006**, *103,* 3357-3362.
47. Lacerda, L.; Soundararajan, A.; Singh, R.; Pastorin, G.; Al-Jamal, K. T.; Turton, J.; Frederik, P.; Herrero, M. A.; Bao, S. L. A.; Emfietzoglou, D.; et al. Dynamic Imaging of functionalized multi-walled carbon nanotube systemic circulation and urinary excretion. *Adv. Mater.* **2008**, *20,* 225-230.
48. Yang, S. T.; Guo, W.; Lin, Y.; Deng, X. Y.; Wang, H. F.; Sun, H. F.; Liu, Y. F.; Wang, X.; Wang, W.; Chen, M.; et al. Biodistribution of pristine single-walled carbon nanotubes in vivo. *J. Phys. Chem. C* **2007**, *111,* 17761-17764.
49. Deng, X. Y.; Yang, S. T.; Nie, H. Y.; Wang, H. F. and Liu, Y. F. A generally adoptable radiotracing method for tracking carbon nanotubes in animals. *Nanotechnology* **2008**, *19,* 075101.
50. Cherukuri, P.; Gannon, C. J.; Leeuw, T. K.; Schmidt, H. K.; Smalley, R. E.; Curley, S. A. and Weisman, R. B. Mammalian pharmacokinetics of carbon nanotubes using intrinsic near-infrared fluorescence. *Proc. Natl. Acad. Sci. U. S. A.* **2006**, *103,* 18882-18886.
51. Liu, Z.; Winters, M.; Holodniy, M. and Dai, H. J. siRNA delivery into human T cells and primary cells with carbon-nanotube transporters. *Angew. Chem. Int. Ed.* **2007**, *46,* 2023-2027.
52. Wu, W.; Wieckowski, S.; Pastorin, G.; Benincasa, M.; Klumpp, C.; Briand, J. P.; Gennaro, R.; Prato, M. and Bianco, A. Targeted delivery of amphotericin B to cells by using functionalized carbon nanotubes. *Angew. Chem. Int. Ed.* **2005**, *44,* 6358-6362.
53. Dumortier, H.; Lacotte, S.; Pastorin, G.; Marega, R.; Wu, W.; Bonifazi, D.; Briand, J. P.; Prato, M.; Muller, S. and Bianco, A. Functionalized carbon nanotubes are non-cytotoxic and preserve the functionality of primary immune cells. *Nano Lett.* **2006**, *6,* 3003-3003.
54. Chen, X.; Lee, G. S.; Zettl, A. and Bertozzi, C. R. Biomimetic engineering of carbon nanotubes by using cell surface mucin mimics. *Angew. Chem. Int. Ed.* **2004**, *43,* 6111-6116.
55. Chen, X.; Tam, U. C.; Czlapinski, J. L.; Lee, G. S.; Rabuka, D.; Zettl, A. and Bertozzi, C. R. Interfacing carbon nanotubes with living cells. *J. Am. Chem. Soc.* **2006**, *128,* 6292-6293.
56. Chin, S. F.; Baughman, R. H.; Dalton, A. B.; Dieckmann, G. R.; Draper, R. K.; Mikoryak, C.; Musselman, I. H.; Poenitzsch, V. Z.; Xie, H. and Pantano, P. Amphiphilic helical peptide enhances the uptake of single-walled carbon nanotubes by living cells. *Exper. Biol. Med.* **2007**, *232,* 1236-1244.
57. Yehia, H. N.; Draper, R. K.; Mikoryak, C.; Walker, E. K.; Bajaj, P.; Musselman, I. H.; Daigrepont, M. C.; Dieckmann, G. R. and Pantano1, P. Single-walled carbon nanotube interactions with HeLa cells. *J. Nanobiotech.* **2007**, *5,* 8.
58. Schipper, M. L.; Nakayama-Ratchford, N.; Davis, C. R.; Kam, N. W. S.; Chu, P.; Liu, Z.; Sun, X.; Dai, H. and Gambhir, S. S. A pilot toxicology study of single-walled carbon nanotubes in a small sample of mice. *Nat. Nanotech.* **2008**, *3,* 216 - 221.
59. Niyogi, S.; Hamon, M. A.; Hu, H.; Zhao, B.; Bhowmik, P.; Sen, R.; Itkis, M. E. and Haddon, R. C. Chemistry of single-walled carbon nanotubes. *Acc. Chem. Res.* **2002**, *35,* 1105-1113.
60. Rosca, I. D.; Watari, F.; Uo, M. and Akaska, T. Oxidation of multiwalled carbon nanotubes by nitric acid. *Carbon* **2005**, *43,* 3124-3131.
61. Zeng, L.; Alemany, L. B.; Edwards, C. L. and Barron, A. R. Demonstration of Covalent Sidewall Functionalization of Single Wall Carbon Nanotubes by NMR Spectroscopy: Side Chain Length Dependence on the Observation of the Sidewall sp3 Carbons. *Nano Res.* **2008**, *1,* 72-88.
62. Zhao, B.; Hu, H.; Yu, A. P.; Perea, D. and Haddon, R. C. Synthesis and characterization of water soluble single-walled carbon nanotube graft copolymers. *J. Am. Chem. Soc.* **2005**, *127,* 8197-8203.
63. Lee, K. M.; Li, L. C. and Dai, L. M. Asymmetric end-functionalization of multi-walled carbon nanotubes. *J. Am. Chem. Soc.* **2005**, *127,* 4122-4123.
64. Moghaddam, M. J.; Taylor, S.; Gao, M.; Huang, S. M.; Dai, L. M. and McCall, M. J. Highly efficient binding of DNA on the sidewalls and tips of carbon nanotubes using photochemistry. *Nano Lett.* **2004**, *4,* 89-93.
65. Coleman, K. S.; Bailey, S. R.; Fogden, S. and Green, M. L. H. Functionalization of single-walled carbon nanotubes via the Bingel reaction. *J. Am. Chem. Soc.* **2003**, *125,* 8722-8723.
66. Umeyama, T.; Tezuka, N.; Fujita, M.; Matano, Y.; Takeda, N.; Murakoshi, K.; Yoshida, K.; Isoda, S. and Imahori, H. Retention of intrinsic electronic properties of soluble single-walled carbon nanotubes after a significant degree of sidewall functionalization by the bingel reaction. *J. Phys. Chem. C* **2007**, *111,* 9734-9741.
67. Georgakilas, V.; Kordatos, K.; Prato, M.; Guldi, D. M.; Holzinger, M. and Hirsch, A. Organic functionalization of carbon nanotubes. *J. Am. Chem. Soc.* **2002**, *124,* 760-761.
68. Tagmatarchis, N. and Prato, M. Functionalization of carbon nanotubes via 1,3-dipolar cycloadditions. *J. Mater. Chem.* **2004**, *14,* 437-439.
69. Pastorin, G.; Wu, W.; Wieckowski, S.; Briand, J. P.; Kostarelos, K.; Prato, M. and Bianco, A. Double functionalisation of carbon nanotubes for multimodal drug delivery. *Chem. Commun.* **2006**, 1182-1184.
70. Chen, R. J.; Zhang, Y. G.; Wang, D. W. and Dai, H. J. Noncovalent sidewall functionalization of single-walled carbon nanotubes for protein immobilization. *J. Am. Chem. Soc.* **2001**, *123,* 3838-3839.
71. Chen, J.; Liu, H. Y.; Weimer, W. A.; Halls, M. D.; Waldeck, D. H. and Walker, G. C. Noncovalent engineering of carbon nanotube surfaces by rigid, functional conjugated polymers. *J. Am. Chem. Soc.* **2002**, *124,* 9034-9035.
72. Wu, P.; Chen, X.; Hu, N.; Tam, U. C.; Blixt, O.; Zettl, A. and Bertozzi, C. R. Biocompatible carbon nanotubes generated by functionalization with glycodendrimers. *Angew. Chem. Int. Ed.* **2008**, *47,* 5022-5025.
73. Zheng, M.; Jagota, A.; Semke, E. D.; Diner, B. A.; Mclean, R. S.; Lustig, S. R.; Richardson, R. E. and Tassi, N. G. DNA-assisted dispersion and separation of carbon nanotubes. *Nat. Mater.* **2003**, *2,* 338-342.

74. Tu, X. and Zheng, M. A DNA-based approach to the carbon nanotube sorting problem. *Nano Res.* **2008**, *1,* 185-194.
75. Moon, H. K.; Chang, C. I.; Lee, D.-K. and Choi, H. C. Effect of nucleases on the cellular internalization of fluorescent labeled DNA-functionalized single-walled carbon nanotubes. *Nano Res.* **2008**, *1,* 351-360.
76. Nakayama-Ratchford, N.; Bangsaruntip, S.; Sun, X. M.; Welsher, K. and Dai, H. J. Noncovalent functionalization of carbon nanotubes by fluorescein-polyethylene glycol: Supramolecular conjugates with pH-dependent absorbance and fluorescence. *J. Am. Chem. Soc.* **2007**, *129,* 2448-2449.
77. Guldi, D. M.; Taieb, H.; Rahman, G. M. A.; Tagmatarchis, N. and Prato, M. Novel photoactive single-walled carbon nanotube-porphyrin polymer wraps: Efficient and long-lived intracomplex charge separation. *Adv. Mater.* **2005**, *17,* 871-875.
78. Richard, C.; Balavoine, F.; Schultz, P.; Ebbesen, T. W. and Mioskowski, C. Supramolecular self-assembly of lipid derivatives on carbon nanotubes. *Science* **2003**, *300,* 775-778.
79. Wang, H.; Zhou, W.; Ho, D. L.; Winey, K. I.; Fischer, J. E.; Glinka, C. J. and Hobbie, E. K. Dispersing single-walled carbon nanotubes with surfactants: A small angle neutron scattering study. *Nano Lett.* **2004**, *4,* 1789-1793.
80. Wong, S. S.; Joselevich, E.; Woolley, A. T.; Cheung, C. L. and Lieber, C. M. Covalently functionalized nanotubes as nanometre-sized probes in chemistry and biology. *Nature* **1998**, *394,* 52-55.
81. Lin, Y.; Taylor, S.; Li, H. P.; Fernando, K. A. S.; Qu, L. W.; Wang, W.; Gu, L. R.; Zhou, B. and Sun, Y. P. Advances toward bioapplications of carbon nanotubes. *J. Mater. Chem.* **2004**, *14,* 527-541.
82. Balavoine, F.; Schultz, P.; Richard, C.; Mallouh, V.; Ebbesen, T. W. and Mioskowski, C. Helical crystallization of proteins on carbon nanotubes: A first step towards the development of new biosensors. *Angew. Chem. Int. Ed.* **1999**, *38,* 1912-1915.
83. Shim, M.; Kam, N.; Chen, R.; Li, Y. and Dai, H. Functionalization of carbon nanotubes for biocompatibility and biomolecular recognition. *Nano Lett.* **2002**, *2,* 285-288.
84. Azamian, B. R.; Davis, J. J.; Coleman, K. S.; Bagshaw, C. B. and Green, M. L. H. Bioelectrochemical single-walled carbon nanotubes. *J. Am. Chem. Soc.* **2002**, *124,* 12664-12665.
85. Erlanger, B. F.; Chen, B. X.; Zhu, M. and Brus, L. Binding of an anti-fullerene IgG monoclonal antibody to single wall carbon nanotubes. *Nano Lett.* **2001**, *1,* 465-467.
86. Wang, S. Q.; Humphreys, E. S.; Chung, S. Y.; Delduco, D. F.; Lustig, S. R.; Wang, H.; Parker, K. N.; Rizzo, N. W.; Subramoney, S.; Chiang, Y. M.; et al. Peptides with selective affinity for carbon nanotubes. *Nat. Mater.* **2003**, *2,* 196-200.
87. Chen, R. J.; Choi, H. C.; Bangsaruntip, S.; Yenilmez, E.; Tang, X. W.; Wang, Q.; Chang, Y. L. and Dai, H. J. An investigation of the mechanisms of electronic sensing of protein adsorption on carbon nanotube devices. *J. Am. Chem. Soc.* **2004**, *126,* 1563-1568.

88. Chandler, D. Interfaces and the driving force of hydrophobic assembly. *Nature* **2005**, *437,* 640-647.
89. Karajanagi, S. S.; Vertegel, A. A.; Kane, R. S. and Dordick, J. S. Structure and function of enzymes adsorbed onto single-walled carbon nanotubes. *Langmuir* **2004**, *20,* 11594-11599.
90. Kong, J.; Franklin, N. R.; Zhou, C. W.; Chapline, M. G.; Peng, S.; Cho, K. J. and Dai, H. J. Nanotube molecular wires as chemical sensors. *Science* **2000**, *287,* 622-625.
91. Byon, H. R. and Choi, H. C. Network single-walled carbon nanotube-field effect transistors (SWNT-FETs) with increased Schottky contact area for highly sensitive biosensor applications. *J. Am. Chem. Soc.* **2006**, *128,* 2188-2189.
92. Kim, S. N.; Rusling, J. F. and Papadimitrakopoulos, F. Carbon nanotubes for electronic and electrochemical detection of biomolecules. *Adv. Mater.* **2007**, *19,* 3214-3228.
93. Wang, J. Carbon-nanotube based electrochemical biosensors: A review. *Electroanalysis* **2005**, *17,* 7-14.
94. Leeuw, T. K.; Reith, R. M.; Simonette, R. A.; Harden, M. E.; Cherukuri, P.; Tsyboulski, D. A.; Beckingham, K. M. and Weisman, R. B. Single-walled carbon nanotubes in the intact organism: Near-IR imaging and biocompatibility studies in Drosophila. *Nano Lett.* **2007**, *7,* 2650-2654.
95. Jin, H.; Heller, D. A. and Strano, M. S. Single-particle tracking of endocytosis and exocytosis of single-walled carbon nanotubes in NIH-3T3 cells. *Nano Lett.* **2008**, *8,* 1577-1585.
96. Barone, P. W.; Parker, R. S. and Strano, M. S. In vivo fluorescence detection of glucose using a single-walled carbon nanotube optical sensor: Design, fluorophore properties, advantages, and disadvantages. *Anal. Chem.* **2005**, *77,* 7556-7562.
97. Jeng, E. S.; Moll, A. E.; Roy, A. C.; Gastala, J. B. and Strano, M. S. Detection of DNA Hybridization Using the Near-Infrared Band-Gap Fluorescence of Single-Walled Carbon Nanotubes. *Nano Lett.* **2006**, *6,* 371-375.
98. Barone, P. W.; Baik, S.; Heller, D. A. and Strano, M. S. Near-infrared optical sensors based on single-walled carbon nanotubes. *Nat. Mater.* **2004**, *4,* 86-92.
99. Satishkumar, B. C.; Brown, L. O.; Gao, Y.; Wang, C. C.; Wang, H. L. and Doorn, S. K. Reversible fluorescence quenching in carbon nanotubes for biomolecular sensing. *Nat. Nanotech.* **2007**, *2,* 560-564.
100. Yang, R. H.; Jin, J. Y.; Chen, Y.; Shao, N.; Kang, H. Z.; Xiao, Z.; Tang, Z. W.; Wu, Y. R.; Zhu, Z. and Tan, W. H. Carbon nanotube-quenched fluorescent oligonucleotides: Probes that fluoresce upon hybridization. *J. Am. Chem. Soc.* **2008**, *130,* 8351-8358.
101. Crochet, J.; Clemens, M. and Hertel, T. Quantum yield heterogeneities of aqueous single-wall carbon nanotube suspensions. *J. Am. Chem. Soc.* **2007**, *129,* 8058-+.
102. Zhang, G. Y.; Qi, P. F.; Wang, X. R.; Lu, Y. R.; Li, X. L.; Tu, R.; Bangsaruntip, S.; Mann, D.; Zhang, L. and Dai, H. J. Selective etching of metallic carbon nanotubes by gas-phase reaction. *Science* **2006**, *314,* 974-977.
103. MacBeath, G. and Schreiber, S. L. Printing proteins as microarrays for high-throughput function determination. *Science* **2000**, *289,* 1760-1763.

104. Bailey, R. C.; Kwong, G. A.; Radu, C. G.; Witte, O. N. and Heath, J. R. DNA-encoded antibody libraries: a unified platform for multiplexed cell sorting and detection of genes and proteins. *J. Am. Chem. Soc.* **2007**, *129,* 1959-1967.
105. Robinson, W. H.; DiGennaro, C.; Hueber, W.; Haab, B. B.; Kamachi, M.; Dean, E. J.; Fournel, S.; Fong, D.; Genovese, M. C.; de Vegvar, H. E.; et al. Autoantigen microarrays for multiplex characterization of autoantibody responses. *Nat. Med.* **2002**, *8,* 295-301.
106. Nie, S. and Emory, S. R. Probing single molecules and single nanoparticles by surface-enhanced Raman scattering. *Science* **1997**, *275,* 1102-1106.
107. Jeanmaire, D. L. and Vanduyne, R. P. Surface Raman Spectroelectrochemistry.1. Heterocyclic, Aromatic, and Aliphatic-Amines Adsorbed on Anodized Silver Electrode. *J. Electroanal. Chem.* **1977**, *84,* 1-20.
108. Espina, V.; Woodhouse, E. C.; Wulfkuhle, J.; Asmussen, H. D.; Petricoin, E. F., 3rd and Liotta, L. A. Protein microarray detection strategies: focus on direct detection technologies. *J. Immunol. Methods* **2004**, *290,* 121-133.
109. Prakash, A.; Mallick, P.; Whiteaker, J.; Zhang, H.; Paulovich, A.; Flory, M.; Lee, H.; Aebersold, R. and Schwikowski, B. Signal maps for mass spectrometry-based comparative proteomics. *Mol. Cell. Proteomics* **2006**, *5,* 423-432.
110. Shvedova, A. A.; Kisin, E. R.; Mercer, R.; Murray, A. R.; Johnson, V. J.; Potapovich, A. I.; Tyurina, Y. Y.; Gorelik, O.; Arepalli, S.; Schwegler-Berry, D.; et al. Unusual inflammatory and fibrogenic pulmonary responses to single-walled carbon nanotubes in mice. *Am. J. Phys. Lung Cell. Mol. Physiol.* **2005**, *289,* L698-L708.
111. Muller, J.; Huaux, F.; Moreau, N.; Misson, P.; Heilier, J. F.; Delos, M.; Arras, M.; Fonseca, A.; Nagy, J. B. and Lison, D. Respiratory toxicity of multi-wall carbon nanotubes. *Toxicol. Appl. Pharmacol.* **2005**, *207,* 221-231.
112. Sayes, C. M.; Liang, F.; Hudson, J. L.; Mendez, J.; Guo, W. H.; Beach, J. M.; Moore, V. C.; Doyle, C. D.; West, J. L.; Billups, W. E.; et al. Functionalization density dependence of single-walled carbon nanotubes cytotoxicity in vitro. *Toxicol. Lett.* **2006**, *161,* 135-142.
113. Dong, L.; Joseph, K. L.; Witkowski, C. M. and Craig, M. M. Cytotoxicity of single-walled carbon nanotubes suspended in various surfactants. *Nanotechnology* **2008**, *19,* 255702.
114. Plata, D. L.; Gschwend, P. M. and Reddy, C. M. Industrially synthesized single-walled carbon nanotubes: compositional data for users, environmental risk assessments, and source apportionment. *Nanotechnology* **2008**, *19,* -.
115. Casey, A.; Herzog, E.; Davoren, M.; Lyng, F. M.; Byrne, H. J. and Chambers, G. Spectroscopic analysis confirms the interactions between single walled carbon nanotubes and various dyes commonly used to assess cytotoxicity. *Carbon* **2007**, *45,* 1425-1432.
116. Worle-Knirsch, J. M.; Pulskamp, K. and Krug, H. F. Oops they did it again! Carbon nanotubes hoax scientists in viability assays. *Nano Lett.* **2006**, *6,* 1261-1268.
117. Yang, S. T.; Wang, X.; Jia, G.; Gu, Y.; Wang, T.; Nie, H.; Ge, C.; Wang, H. and Liu, Y. Long-term accumulation and low toxicity of single-walled carbon nanotubes in intravenously exposed mice. *Toxicol. Lett.* **2008**, *181,* 182-189.

118. Yang, S. T.; Fernando, K. A.; Liu, J. H.; Wang, J.; Sun, H. F.; Liu, Y.; Chen, M.; Huang, Y.; Wang, X.; Wang, H.; et al. Covalently PEGylated carbon nanotubes with stealth character in vivo. *Small* **2008**, *4,* 940-944.
119. Salvador-Morales, C.; Flahaut, E.; Sim, E.; Sloan, J.; Green, M. L. H. and Sim, R. B. Complement activation and protein adsorption by carbon nanotubes. *Mol. Immunol.* **2006**, *43,* 193-201.
120. Kostarelos, K.; Lacerda, L.; Pastorin, G.; Wu, W.; Wieckowski, S.; Luangsivilay, J.; Godefroy, S.; Pantarotto, D.; Briand, J. P.; Muller, S.; et al. Cellular uptake of functionalized carbon nanotubes is independent of functional group and cell type. *Nat. Nanotech.* **2007**, *2,* 108-113.
121. Feazell, R. P.; Nakayama-Ratchford, N.; Dai, H. and Lippard, S. J. Soluble single-walled carbon nanotubes as longboat delivery systems for Platinum(IV) anticancer drug design. *J. Am. Chem. Soc.* **2007**, *129,* 8438-+.
122. Ali-Boucetta, H.; Al-Jamal, K. T.; McCarthy, D.; Prato, M.; Bianco, A. and Kostarelos, K. Multiwalled carbon nanotube-doxorubicin supramolecular complexes for cancer therapeutics. *Chem. Commun.* **2008**, 459-461.
123. Murakami, T.; Fan, J.; Yudasaka, M.; Iijima, S. and Shiba, K. Solubilization of single-wall carbon nanohorns using a PEG-doxorubicin conjugate. *Mol. Pharmaceutics* **2006**, *3,* 407 -414.
124. Sun, X.; Liu, Z.; Welsher, K.; Robinson, J. T.; Goodwin, A.; Zaric, S. and Dai, H. Nano-graphene oxide for cellular imaging and drug delivery. *Nano Res.* **2008**, *1,* 203-212.
125. Liu, Z.; Robinson, J. T.; Sun, X. M. and Dai, H. J. PEGylated nanographene oxide for delivery of water-insoluble cancer drugs. *J. Am. Chem. Soc.* **2008**, *130,* 10876-10877.
126. Dhar, S.; Liu, Z.; Thomale, J.; Dai, H. and Lippard, S. J. Targeted Single-Wall Carbon Nanotube-Mediated Pt(IV) Prodrug Delivery Using Folate as a Homing Device. *J. Am. Chem. Soc.* **2008**, *130,* 11467–11476.
127. Liu, Z.; Li, X.; Tabakman, S. M.; Jiang, K.; Fan, S. and Dai, H. Multiplexed multi-color Raman imaging of live cells with isotopically modified single walled carbon nanotubes. *J. Am. Chem. Soc.* **2008**, *130,* 13540–13541.
128. Chakravarty, P.; Marches, R.; Zimmerman, N. S.; Swafford, A. D.; Bajaj, P.; Musselman, I. H.; Pantano, P.; Draper, R. K. and Vitetta, E. S. Thermal ablation of tumor cells with antibody-functionalized single-walled carbon nanotubes. *Proc. Natl. Acad. Sci. U. S. A.* **2008**, *105,* 8697-8702.
129. McDevitt, M. R.; Chattopadhyay, D.; Kappel, B. J.; Jaggi, J. S.; Schiffman, S. R.; Antczak, C.; Njardarson, J. T.; Brentjens, R. and Scheinberg, D. A. Tumor targeting with antibody-functionalized, radiolabeled carbon nanotubes. *J. Nucl. Med.* **2007**, *48,* 1180-1189.
130. Kataura, H.; Maniwa, Y.; Kodama, T.; Kikuchi, K.; Hirahara, K.; Suenaga, K.; Iijima, S.; Suzuki, S.; Achiba, Y. and Kratschmer, W. High-yield fullerene encapsulation in single-wall carbon nanotubes. *Synth. Met.* **2001**, *121,* 1195-1196.
131. Jeong, G. H.; Farajian, A. A.; Hatakeyama, R.; Hirata, T.; Yaguchi, T.; Tohji, K.; Mizuseki, H. and Kawazoe, Y. Cesium encapsulation in single-walled carbon nanotubes via plasma ion irradiation: Application to junction formation and ab initio investigation. *Phys. Rev. B* **2003**, *68,* 075410.

132. Li, L. J.; Khlobystov, A. N.; Wiltshire, J. G.; Briggs, G. A. D. and Nicholas, R. J. Diameter-selective encapsulation of metallocenes in single-walled carbon nanotubes. *Nat. Mater.* **2005**, *4,* 481-485.
133. Kaneko, T.; Okada, T. and Hatakeyama, R. DNA encapsulation inside carbon nanotubes using micro electrolyte plasmas. *Contrib. Plasma Phys.* **2007**, *47,* 57-63.
134. Hilder, T. A. and Hill, J. M. Modelling the encapsulation of the anticancer drug cisplatin into carbon nanotubes. *Nanotechnology* **2007**, *18,* -.
135. Hilder, T. A. and Hill, J. M. Probability of encapsulation of paclitaxel and doxorubicin into carbon nanotubes. *Micro Nano Lett.* **2008**, *3,* 41-49.
136. Mello, C. C. and Conte, D. Revealing the world of RNA interference. *Nature* **2004**, *431,* 338-342.
137. Marshall, E. Clinical trials - Gene therapy death prompts review of adenovirus vector. *Science* **1999**, *286,* 2244-2245.
138. Hacein-Bey-Abina, S.; von Kalle, C.; Schmidt, M.; Le Deist, F.; Wulffraat, N.; McIntyre, E.; Radford, I.; Villeval, J. L.; Fraser, C. C.; Cavazzana-Calvo, M.; et al. A serious adverse event after successful gene therapy for X-linked severe combined immunodeficiency. *New Engl. J. Med.* **2003**, *348,* 255-256.
139. Zhang, Z. H.; Yang, X. Y.; Zhang, Y.; Zeng, B.; Wang, Z. J.; Zhu, T. H.; Roden, R. B. S.; Chen, Y. S. and Yang, R. C. Delivery of telomerase reverse transcriptase small interfering RNA in complex with positively charged single-walled carbon nanotubes suppresses tumor growth. *Clin. Cancer Res.* **2006**, *12,* 4933-4939.
140. Choi, H. S.; Liu, W.; Misra, P.; Tanaka, E.; Zimmer, J. P.; Ipe, B. I.; Bawendi, M. G. and Frangioni, J. V. Renal clearance of quantum dots. *Nat. Biotech.* **2007**, *25,* 1165-1170.
141. Wang, H. F.; Wang, J.; Deng, X. Y.; Sun, H. F.; Shi, Z. J.; Gu, Z. N.; Liu, Y. F. and Zhao, Y. L. Biodistribution of carbon single-wall carbon nanotubes in mice. *J. Nanosci. Nanotech.* **2004**, *4,* 1019-1024.
142. Deng, X.; Jia, G.; Wang, H.; Sun, H.; Wang, X.; Yang, S.; Wang, T. and Liu, Y. Translocation and fate of multi-walled carbon nanotubes in vivo. *Carbon* **2007**, *45,* 1419-1424.
143. Pavlinkova, G.; Beresford, G. W.; Booth, B. J. M.; Batra, S. K. and Colcher, D. Pharmacokinetics and biodistribution of engineered single-chain antibody constructs of MAb CC49 in colon carcinoma xenografts. *J. Nucl. Med.* **1999**, *40,* 1536-1546.
144. Olmsted, S. S.; Padgett, J. L.; Yudin, A. I.; Whaley, K. J.; Moench, T. R. and Cone, R. A. Diffusion of macromolecules and virus-like particles in human cervical mucus. *Biophys. J.* **2001**, *81,* 1930-1937.
145. Goel, A.; Colcher, D.; Baranowska-Kortylewicz, J.; Augustine, S.; Booth, B. J. M.; Pavlinkova, G. and Batra, S. K. Genetically engineered tetravalent single-chain Fv of the pancarcinoma monoclonal antibody CC49: Improved biodistribution and potential for therapeutic application. *Cancer Res.* **2000**, *60,* 6964-6971.
146. Aubin, J. E. Autofluorescence of Viable Cultured Mammalian-Cells. *J. Histochem. Cytochem.* **1979**, *27,* 36-43.

147. Wang, F.; Dukovic, G.; Brus, L. E. and Heinz, T. F. Time-resolved fluorescence of carbon nanotubes and its implication for radiative lifetimes. *Phys. Rev. Lett.* **2004**, *92,* 177401.
148. Lefebvre, J.; Austing, D. G.; Bond, J. and Finnie, P. Photoluminescence imaging of suspended single-walled carbon nanotubes. *Nano Lett.* **2006**, *6,* 1603-1608.
149. Reich, S.; Dworzak, M.; Hoffmann, A.; Thomsen, C. and Strano, M. S. Excited-state carrier lifetime in single-walled carbon nanotubes. *Phys. Rev. B* **2005**, *71,* 033402.
150. Cognet, L.; Tsyboulski, D. A.; Rocha, J. D. R.; Doyle, C. D.; Tour, J. M. and Weisman, R. B. Stepwise quenching of exciton fluorescence in carbon nanotubes by single-molecule reactions. *Science* **2007**, *316,* 1465-1468.
151. Heller, D. A.; Mayrhofer, R. M.; Baik, S.; Grinkova, Y. V.; Usrey, M. L. and Strano, M. S. Concomitant length and diameter separation of single-walled carbon nanotubes. *J. Am. Chem. Soc.* **2004**, *126,* 14567-14573.
152. Crochet, J.; Clemens, M. and Hertel, T. Quantum yield heterogeneities of aqueous single-wall carbon nanotube suspensions. *J. Am. Chem. Soc.* **2007**, *129,* 8058-8059.
153. Keren, S.; Zavaleta, C.; Cheng, Z.; de la Zerda, A.; Gheysens, O. and Gambhir, S. S. Noninvasive molecular imaging of small living subjects using Raman spectroscopy. *Proc. Nat. Acad. Sci. U. S. A.* **2008**, *105,* 5844-5849.
154. Zavaleta, C.; Zerda, A. d. l.; Liu, Z.; Keren, S.; Cheng, Z.; Schipper, M.; Chen, X.; Dai, H. and Gambhir, S. S. Noninvasive Raman Spectroscopy in Living Mice for Evaluation of Tumor Targeting with Carbon Nanotubes. *Nano Lett.* **2008**, *8,* 2800–2805.
155. Xu, M. H. and Wang, L. H. V. Photoacoustic imaging in biomedicine. *Rev. Sci. Instrum.* **2006**, *77,* 041101.
156. Sun, X.; Zaric, S.; Daranciang, D.; Welsher, K.; Lu, Y.; Li, X. and Dai, H. Optical properties of ultrashort semiconducting single-walled carbon nanotube capsules down to sub-10 nm. *J. Am. Chem. Soc.* **2008**, *130,* 6551-6555.
157. Li, X. L.; Tu, X. M.; Zaric, S.; Welsher, K.; Seo, W. S.; Zhao, W. and Dai, H. J. Selective synthesis combined with chemical separation of single-walled carbon nanotubes for chirality selection. *J. Am. Chem. Soc.* **2007**, *129,* 15770-15771.
158. Jorio, A.; Saito, R.; Hafner, J. H.; Lieber, C. M.; Hunter, M.; McClure, T.; Dresselhaus, G. and Dresselhaus, M. S. Structural (n, m) determination of isolated single-wall carbon nanotubes by resonant Raman scattering. *Phys. Rev. Lett.* **2001**, *86,* 1118-1121.